\documentclass[lettersize,journal]{IEEEtran}
\usepackage{amsmath,amsfonts}
\usepackage{algorithmic}
\usepackage{algorithm}
\usepackage{array}
\usepackage[caption=false,font=normalsize,labelfont=sf,textfont=sf]{subfig}
\usepackage{textcomp}
\usepackage{stfloats}
\usepackage{url}
\usepackage{verbatim}
\usepackage{graphicx}
\usepackage{cite}
\usepackage{hyperref}
\usepackage[nameinlink]{cleveref}
\usepackage{xcolor}
\usepackage{enumitem}
\newcommand{\aka}{{a.k.a.}}
\newcommand{\ie}{\textit{i.e.}}
\newcommand{\etal}{\textit{et al.}}
\newcommand{\eg}{\textit{e.g.}}
\newcommand{\para}[1]{\phantomsection\vspace{1mm}\noindent\textbf{#1}}

\newcommand{\userquote}[1]{\textit{``#1''}}

\newcommand{\re}[2]{{\color{black} #1}}

\newcommand{\retvcg}[2]{{\color{black} #1}}

\AtBeginDocument{

}

\AtBeginDocument{%
  \mathchardef\mathcomma\mathcode`\,
  \mathcode`\,="8000 
}
{\catcode`,=\active
  \gdef,{\mathcomma\discretionary{}{}{}}
}

\begin{document}

\title{Exploring Spatial Hybrid User Interface\\ for Visual Sensemaking}

\author{Wai~Tong,
    Haobo~Li,
    Meng~Xia,
    Wong~Kam-Kwai,
    Ting-Chuen~Pong,
    Huamin~Qu,
    and~Yalong~Yang
\thanks{Wai Tong is with Texas
A\&M University, Texas, USA. }
\thanks{Haobo Li is with Hong Kong University of Science and Technology, Hong Kong, China.}
\thanks{Meng Xia is with Texas
A\&M University, Texas, USA.}
\thanks{Wong Kam-Kwai is with Hong Kong University of Science and Technology, Hong Kong, China.}
\thanks{Ting-Chuen Pong is with Hong Kong University of Science and Technology, Hong Kong, China.}
\thanks{Huamin Qu is with Hong Kong University of Science and Technology, Hong Kong, China.}
\thanks{Yalong Yang is with Georgia Institute of Technology, Georgia, USA. He is the corresponding author.}
\thanks{Manuscript received April 19, 2021; revised August 16, 2021.}}

\markboth{Journal of \LaTeX\ Class Files,~Vol.~14, No.~8, August~2021}%
{Shell \MakeLowercase{\textit{et al.}}: A Sample Article Using IEEEtran.cls for IEEE Journals}

\IEEEpubid{0000--0000/00\$00.00~\copyright~2021 IEEE}

\maketitle

\begin{abstract}
We built a spatial hybrid system that combines a personal computer (PC) and virtual reality (VR) for visual sensemaking, addressing limitations in both environments.
Although VR offers immense potential for interactive data visualization (e.g., large display space and spatial navigation), it can also present challenges such as imprecise interactions and user fatigue. At the same time, a PC offers precise and familiar interactions but has limited display space and interaction modality. Therefore, we iteratively designed a spatial hybrid system (PC+VR) to complement these two environments by enabling seamless switching between PC and VR environments.
To evaluate the system's effectiveness and user experience, we compared it to using a single computing environment (\ie{}, PC-only and VR-only).
Our study results (N=18) showed that spatial PC+VR could combine the benefits of both devices to outperform user preference for VR-only without a negative impact on performance from device switching overhead. 
Finally, we discussed future design implications.
\end{abstract}


\section{Introduction}
\label{sec:intro}

\IEEEPARstart{S}{ensemaking} \retvcg{is the process of interpreting and understanding complex information or situations to create insights and inform decision-making~\cite{pirolli2005sensemaking}}{}.
With increasingly available data, data-driven sensemaking becomes ubiquitous.
Visual user interfaces and data visualization have been shown to be helpful and effective for data-driven sensemaking,
as they augment people's ability to recognize patterns and distill insights from a large and complex dataset~\cite{card1999readings,andrews2010space,endert2012semantic}.
\retvcg{We refer to this process of making sense of data through visual user interfaces and data visualization as \textbf{visual sensemaking}.}{}
To support visual sensemaking, displays are essential to represent the data visually, and interactions are crucial to manipulate the data and visualizations to match the user's mental model.

\retvcg{While many visual sensemaking applications are currently tailored for the desktop environment (or PC), such an environment may be limited in addressing the increasing complexities of real-world problems, constrained by their limited display size, mobility, and interaction modalities. Therefore, a range of computing environments beyond the traditional desktop environment~\cite{munzner2014visualization} have been studied. For example, such as display walls~\cite{ball2007move,belkacem2022interactive} and tabletops~\cite{ens2020uplift} for the larger display size, smartphones for situated analysis~\cite{lee2021mobile}, and smartwatches~\cite{horak2018david} for on-body input capability.}{}
The recent emergence of affordable virtual and augmented reality (VR/AR) head-mounted displays (HMDs) has sparked a growing interest in using VR/AR for interactive data visualization and visual sensemaking.
This has given rise to a new research field known as \textit{Immersive Analytics}~\cite{ens2021grand,marriott2018immersive}.
Immersive analytics offers a range of potential benefits over the traditional desktop environment, such as the ability to use large display spaces~\cite{satriadi2020maps,lisle2020evaluating,yang2020embodied}, embodied and tangible interaction~\cite{yang2020tilt,cordeil2017imaxes,tong2022exploring,in2023table,hurter2018fiberclay}, 3D rendering~\cite{bach2017hologram,yang2018maps,kraus2019impact,kwon2016study,brath20143d}\re{, and spatial navigation~\cite{lisle2020evaluating,in2024evaluating,yang2020embodied,ball2007move,hayatpur2020datahop,li2023gestureexplorer}}{}.

However, immersive analytics has limitations. For example, users have reported experiencing fatigue when using VR/AR systems~\cite{mcmahan2012evaluating}. Additionally, performing precise interactions in VR/AR can be challenging, such as inputting specific values or adjusting a range on a slider~\cite{cordeil2020embodied}. 
While efforts have been made to improve the input experience in VR/AR, it remains time-consuming and frustrating in some scenarios.
Conversely, PC, as the most widely used computing environment, excels in tasks where VR/AR falls short, such as text input~\cite{mcgill2015dose}.

\begin{figure}
\centering
\includegraphics[width=\columnwidth]{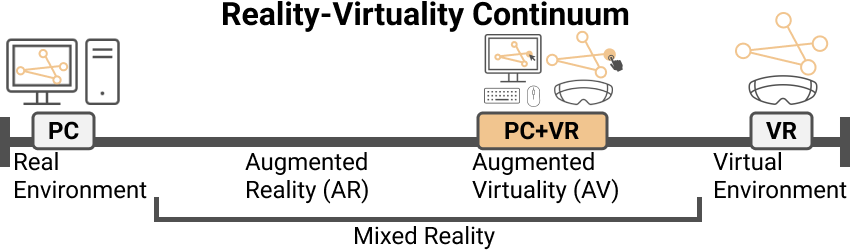}
\caption{The figure shows the PC, our proposed PC+VR hybrid system, and VR in reality–virtuality continuum~\cite{milgram1995augmented}. As the resulting environment is mainly virtual, it leans towards VR and falls under Augmented Virtuality.}
\label{fig:rv}
\end{figure}

\IEEEpubidadjcol %

While both PC and VR platforms offer distinct advantages and present specific challenges, we wanted to explore the concept of a hybrid user interface~\cite{feiner1991hybrid}, aiming to capitalize on their strengths and mitigate their inherent limitations. 
\re{Current hybrid approaches often adopt a transitive approach, which requires users to frequently switch between devices~\cite{hubenschmid2022relive,jansen2023autovis}.
Other hybrid methods~\cite{immersed2023, wang2020towards, pavanatto2021we, wang2022understanding, seraji2024analyzing} utilize non-transitional approaches, which favor a seated posture suited to the desktop environment but restrict spatial navigation within immersive spaces. 
Spatial ability has been shown to be beneficial in sensemaking and data exploration~\cite{lisle2020evaluating,in2024evaluating}.}{}
Therefore, we aim to explore \textbf{how people would utilize a \textit{spatial hybrid PC+VR system} for visual sensemaking}.

\begin{figure}
\centering
\includegraphics[width=\columnwidth]{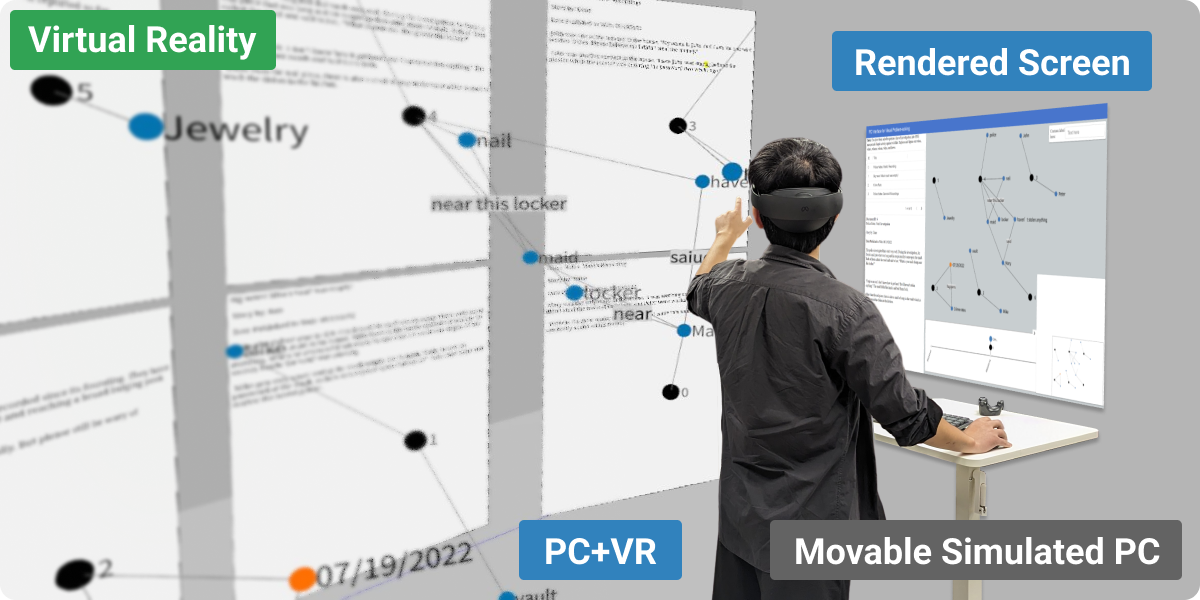}
\caption{A demonstration of spatial hybrid systems for visual problem-solving: users can read documents and build a node-link diagram in VR. They interact with a digitally rendered flat screen on a physically movable table, using a mouse and keyboard to input text annotations. Notably, users do not need to put on and off the VR headset to switch between environments.}
\label{fig:teaser}
\end{figure}

To answer this question, we adopted an iterative design methodology, given the limited existing guidelines on designing spatial hybrid AR/VR systems~\cite{krauss2021current}, particularly in the context of visual sensemaking.
First, based on existing literature (\eg{},~\cite{lisle2020evaluating,hubenschmid2021towards,davidson2022exploring,tong2023towards}), we derived five design requirements: supporting a movable spatial hybrid system, designing optimized interfaces for both PC and VR, providing the same context in both interfaces, allowing non-transitional usage of PC and VR interfaces, and allowing easy-to-switch input modality and cross-device interaction.
\retvcg{Second, we designed a prototype to blend PC and VR interfaces using Augmented Virtuality~\cite{milgram1995augmented} (\Cref{fig:rv}).
As shown in \Cref{fig:teaser}, users can use a simulated PC rendered in a 3D room-sized space in VR. The simulated PC screen's position and rotation are controlled by a tracker placed on a wheeled table so that users can move the virtual screen by physically moving the table. Users can interact with the virtual screen using a mouse and keyboard, creating a similar experience to using a physical ``PC.''}{}
Moreover, the state of the graphs is shared between devices to support cross-device linking and brushing. Lastly, we used hand gestures as the major control for VR because it is easier for users to switch from mouse and keyboard to hand gestures than controllers. Third, we conducted a pilot study with 12 participants to iteratively improve the design and implementation of the spatial hybrid PC+VR system.

\retvcg{In this project, we aim to explore how our proposed spatial hybrid PC+VR system changes user analytics behavior and experience compared to using a single environment, thereby enriching the empirical understanding of spatial hybrid PC+VR systems, particularly in the context of visual sensemaking. 
To achieve this, we conducted a controlled user study with 18 participants, comparing the PC+VR system to two baseline conditions: PC-only and VR-only.}{}
Participants were asked to complete a sensemaking task derived from the literature~\cite{mahyar2014supporting,balakrishnan2008visualizations,tong2023towards,yang2024putting}, requiring them to build a node-link diagram from text documents to answer analytical questions.
We found that the spatial hybrid PC+VR system did not negatively impact task performance, even with the addition of an extra wheeled table compared to previous work~\cite{pavanatto2021we}. Furthermore, it was preferred and reduced physical demand compared to the VR-only system.
We further found four different patterns in both spatial and temporal analysis of users' usage patterns. These findings could provide insight into designing future spatial hybrid PC+VR systems for visual sensemaking.

In summary, our contributions are three-fold.
\textbf{First}, we compiled design requirements for spatial hybrid PC+VR systems for visual sensemaking. 
\textbf{Second}, we iteratively developed a spatial hybrid PC+VR prototype based on these design considerations. 
\textbf{Finally}, we conducted a \retvcg{}{controlled} user study to investigate our system's user experience and performance, which provides insight into future designs.

\section{Related Work}

\para{Data Visualization for Visual Sensemaking.}
Data visualization is shown to be effective for sensemaking by visualizing the complex relationships among entities~\cite{card1999readings}. For example, Hendrix~\etal{}~\cite{hendrix2000visualizations} visualizes the hierarchical structure of Java code for better program comprehension. Perer~\etal{}~\cite{perer2011visual} utilized a node-link diagram to understand the complex relationships between different enterprises. Moreover, researchers~\cite{balakrishnan2008visualizations, lin2021taxthemis, mahyar2014supporting} implemented visualization systems to assist in solving crime.

However, the abovementioned systems are designed for the desktop environment, which provide only a limited 2D display space. Recently, Lee~\etal{}~\cite{lee2021post} tried to bring Post-it notes with links from reality to VR for ideation, and Tong~\etal{}~\cite{tong2023towards} explored visual sensemaking with node-link diagrams in VR.
In this work, we aim to explore the potential of using AR/VR HMDs for visual sensemaking.

\para{Immersive Analytics.} Immersive analytics
is an emerging research field that uses
immersive technologies, such as VR and AR, to interact with and explore complex data in new and more intuitive ways~\cite{marriott2018immersive,ens2021grand}.
A series of studies have been conducted to explore the benefits and drawbacks of Immersive Analytics, and these have been comprehensively reviewed in recent surveys~\cite{fonnet2021survey,kraus2022immersive,skarbez2019immersive,klein2022immersive}.
While we do not aim to provide a comprehensive list of all such studies, we introduce some representative ones to offer an informative overview.

There are several frequently reported benefits of Immersive analytics from the literature: 3D rendering, large display space, embodied interaction\re{, and spatial navigation}{}.
\emph{Firstly}, VR/AR facilitates the rendering of 3D graphics in any location around the user, making it particularly useful for analyzing data with inherent 3D information~\cite{brath20143d}, such as scientific visualizations~\cite{el2019virtual}. 
The additional space and dimension provided by VR/AR also enable the de-cluttering of visualizations with dense data and facilitate the examination of spatial patterns, as seen in geographic visualizations~\cite{yang2018origin}, scatterplots~\cite{kraus2019impact,bach2017hologram}, and network visualizations~\cite{kwon2016study}.
\emph{Secondly}, VR/AR provides an almost unlimited screen that creates new possibilities for creating novel techniques, such as room-sized visualizations~\cite{yang2018maps,kraus2019impact}. 
\emph{Thirdly}, embodied physical movements can be more intuitive and expressive than the conventional PC user interface~\cite{cordeil2017imaxes,yang2020tilt,liu2023datadancing,tong2022exploring,hurter2018fiberclay}. 
\re{Lastly, physical moving in space and head rotation has proven to be an effective navigation method~\cite{yang2020embodied,ball2007move} and is readily available in immersive environments.
Additionally, the spatial arrangement of content leverages spatial memory and leads to more efficient wayfinding~\cite{yang2020virtual,satriadi2020maps,lisle2020evaluating,hayatpur2020datahop,in2024evaluating,li2023gestureexplorer}.}{}

While Immersive Analytics can offer many benefits, there are significant reservations about using VR/AR.
For example, although numerous efforts have been made to enhance the text input experience in VR/AR, its efficiency and usability still lag behind keyboard input in a desktop environment~\cite{speicher2018selection}.
Additionally, the laser pointer is the most commonly used interaction in VR/AR, allowing users to access distant objects. However, such mid-air interactions unavoidably result in fatigue and imprecise interaction with data points~\cite{cordeil2020embodied}. 
Given these limitations, we propose to address its disadvantages by providing users with a PC computing environment.

\para{\re{Hybrid User Interface with PC and Immersive Technology}{Hybrid User Interface} for Data Visualization.}
Since different devices have unique advantages, the hybrid user interface has been introduced to leverage benefits from different devices~\cite{feiner1991hybrid}. However, building hybrid systems poses inherent and distinct challenges, such as loss of context~\cite{hubenschmid2021towards}. 
Furthermore, interaction design across devices continues to be one of the significant challenges in the field~\cite{ens2021grand}.

\re{One popular hybrid user interface design for data visualization would be combining PC and immersive technologies (AR/VR) because visual analytics systems and visualizations are designed and tailored for a PC environment. On the one hand, researchers explore transitional and collaborative interfaces for data exploration~\cite{frohler2022survey}.}{} 
For example, ReLive has been introduced to provide ex-situ analysis on PC and in-situ analysis in VR~\cite{hubenschmid2022relive}. However, the proposed system requires users to put on and remove the VR HMD to switch between complete reality and virtuality, disrupting workflow continuity. AutoVis~\cite{jansen2023autovis} includes a virtual tablet showing part of the visualization from the PC view. It potentially reduces context switching. Yet, interactions with the desktop view are not fully supported in the VR environment. \retvcg{It still requires putting on and off the VR headset to switch computing environments. Therefore, the cost of transition remains high.}{}

On the other hand, researchers are actively exploring non-transitional interfaces (\eg{}, using AR and PC simultaneously for visual analytics).
For example, Wang~\etal{}~\cite{wang2020towards,wang2022understanding} has explored the combination of AR and PC for 3D scientific visualization. \re{Seraji~\etal{}~\cite{seraji2024analyzing} enables users to transfer data visualizations between AR and PC environments.}{}

\re{Although the involvement of the AR could make the integration of the PC setup easier, the interference of the background context makes the color~\cite{whitlock2020graphical} and text~\cite{zhou2024did} in visualizations harder to design and read. In particular, our tasks do not incorporate real-world context; therefore, instead of using AR, which may introduce background distractions, VR offers a fully immersive visual analytics experience that enhances focus and minimizes distractions~\cite{lisle2023different}}{}.
\re{Furthermore, in the previously mentioned work, AR/VR primarily serves as an extended 3D display for the PC workspace, where the user remains seated and stationary~\cite{wang2022understanding,pavanatto2021we,immersed2023}}{}. This approach does not fully exploit the spatial capabilities offered by the immersive environment. 
\retvcg{Wang~\etal{}~\cite{wang2020towards} investigated the potential of spatial movement in visual exploration and discovered that navigating through data by walking is intuitive. However, in their study, the PC remained stationary, which meant that users could not utilize the computer interface while moving around, leading to interruptions in their workflow.}{} 

\retvcg{As a result, we aim to design a hybrid spatial system that reduces the cost of transition and the interference of the reality background and can fully leverage the benefits of immersive techniques (\eg{}, large display and spatial navigation).}{}
 
\section{Designing Spatial Hybrid Interfaces for Visual Sensemaking}
\label{sec:pc_vr_design}

Following the methodology established by Horak et al.~\cite{horak2018david} for identifying design requirements of hybrid interfaces, we first selected a representative task scenario. We then analyzed the core design characteristics of PC and VR within this context. Based on these characteristics, we defined the design criteria essential for a spatial hybrid user interface tailored to visual sensemaking tasks.

\subsection{Task Domain}
\label{secc:task}
We chose a classic visual sensemaking task from the literature~\cite{mahyar2014supporting,balakrishnan2008visualizations,tong2023towards,yang2024putting} where the user acts as a detective tasked with investigating a hidden illegal activity against wildlife using a set of text documents. 
Similar to a real-world scenario, the detective must extract key entities from documents and construct a node-link diagram connecting different entities with relationships. 
To complete the task, the user must identify the who, what, where, when, how, and why of the event.
We chose this scenario for several reasons: node-link diagrams are familiar to users and require low visualization literacy; they are justified in both 2D and 3D; and the sensemaking task is complex enough to provide a deeper understanding of the visualization system's user experience.

\subsection{PC and VR for Visual Sensemaking}
\label{sec:pc-vr-only-design}
To design the spatial hybrid interface combining PC and VR for visual sensemaking, we first systematically analyzed the characteristics of each platform.

\para{PC}s generally offer visual interfaces with relatively high-resolution input and output~\cite{feiner1991hybrid}. 
In particular, high-resolution input means that the mouse and keyboard provide a mature and accurate interaction mechanism. Moreover, many well-designed systems have utilized high-resolution displays to compact information into a standard PC screen. 
Additionally, 2D visualization used in PC provides users with a familiar visual representation~\cite{riegler2020cross} and visual exploration workflow.
In general, 2D node-link diagrams have been commonly used to reveal relationships between entities in visual sensemaking~\cite{mahyar2014supporting,balakrishnan2008visualizations,lin2021taxthemis,perer2011visual}. 
However, the PC's small physical display space may limit the scalability of the visualization.

\para{VR} can provide the same \retvcg{views}{interfaces} and functionalities as in the PC interface\retvcg{, except input devices}{}. 
Additionally, an optimized VR interface can leverage its display and interaction modalities---such as a large workspace, 3D visualization, embodied interaction, and spatial navigation---for visual sensemaking tasks.
Specifically, the virtually large display area accommodates more visualizations~\cite{horak2018david} and other content, such as documents~\cite{tong2023towards}, enhancing visual content management~\cite{satriadi2020maps,lisle2020evaluating,yang2020virtual}.
Moreover, 3D node-link diagrams have been shown to outperform traditional 2D node-link diagrams~\cite{kwon2016study}.
Additionally, AR/VR introduces more intuitive and novel interactions for data visualization through physical body movements~\cite{cordeil2017imaxes,liu2023datadancing,tong2022exploring,yang2020tilt,in2023table}.
\re{Lastly, spatial ability can be utilized and beneficial to the sensemaking process in VR~\cite{lisle2020evaluating}, as well as data analytics in VR~\cite{hayatpur2020datahop,li2023gestureexplorer,in2024evaluating}.}{}
Nonetheless, the relatively low-resolution output and input capability limit the effectiveness of visual sensemaking~\cite{feiner1991hybrid}. 
In particular, the text is hard to read, and the input interaction in VR is not precise enough.

\subsection{PC+VR Hybrid System for Visual Sensemaking}

Guided by prior studies~\cite{lisle2020evaluating,hubenschmid2022relive,hubenschmid2021towards,davidson2022exploring,tong2023towards} and our specific research objectives concerning crime-solving tasks, we delineate five key design requirements.

\para{R1: Supporting a movable hybrid user interface.}\label{req:r1}
One of the benefits of the immersive environment is the spatial navigation~\cite{lisle2020evaluating,hayatpur2020datahop,li2023gestureexplorer}. However, the current design of hybrid interfaces of PC and immersive devices does not fully utilize this characteristic. 
Most of the work~\cite{immersed2023,wang2020towards} used the AR/VR HMD to provide an extended 3D view for the PC monitor, and users are mainly seated without any navigation in space. 
To fully engage users in ``\textit{Immersive Space to Think}''~\cite{lisle2020evaluating}, supporting spatial navigation in the hybrid system should be considered. 

\para{R2: Design optimized interface for each device.}\label{req:r2}
Each device's design should be optimized to increase user experience~\cite{tong2023towards}.
For example, as demonstrated in previous studies~\cite{davidson2022exploring,lisle2021sensemaking}, the large 3D space in VR should be utilized to facilitate the spatial sensemaking process. Moreover, Saffo~\etal{}~\cite{saffo2023eyes} suggested that abstract data visualizations are better suited for interpretation and interaction on a PC, while natural spatial mapping visualizations are more advantageous in VR. Thus, the design in PC should make use of the existing well-established design, and the design in VR should make good use of the large 3D space.

\para{R3: Provide the same context in both interfaces.}\label{req:r3}
\re{The PC and VR interfaces should support the same context to avoid losing context when switching devices.
People may prefer to perform the task even in a less efficient environment to avoid the trouble caused by switching devices~\cite{hubenschmid2022relive}.
Therefore, previous work often offers users similar or duplicated views in both devices to provide the same context~\cite{hubenschmid2022relive,jansen2023autovis}.
Especially in our work, we are interested in investigating when users choose to use a PC and when to use VR in the PC+VR system and want to ensure the investigation is unbiased.
As a result, we aim to design the spatial hybrid PC+VR system with identical information and functionalities in both environments.}{}

\para{R4: \retvcg{Reduce transition cost between}{} PC and VR interfaces.}\label{req:r4} \retvcg{Frequent transitions between devices could interrupt the sensemaking process and create disorientation~\cite{hubenschmid2022relive}.
Therefore, we consider minimizing the transition time between PC and VR usage during visual sensemaking.}{}
Inspired by Davidson~\etal{}~\cite{davidson2022exploring} and suggested by Hubenschimid~\etal{}~\cite{hubenschmid2022relive}, we aim to render the PC screen inside VR, \ie{}, a simulated PC in VR~\cite{jetter2020vr}, so that the user could view the PC and VR interfaces at the same time\retvcg{, avoiding the change in devices to reduce transition time.}{}
Thus, we expected the hybrid interface to be located near Augmented Virtuality in reality–virtuality continuum~\cite{milgram1995augmented} (\Cref{fig:rv}), meaning that users are situated in VR but they can still see partial objects in reality, \ie{}, the physical table, keyboard, and mouse.

\para{R5: Allow easy-to-switch input modality and cross-device interaction.} \label{req:r5}
Tedious switching between different input devices could potentially increase the cost of using different devices. For example, although VR controllers could provide more precise control and more functionality compared to hand gestures, they did not complement well with the mouse and keyboard. Users are required to find an empty space to put down the VR controllers whenever they switch to a PC. Therefore, the hybrid system should allow an easy-to-switch input modality, such as hand.
Moreover, cross-device linking and brushing should be supported~\cite{hubenschmid2022relive}. Users should be able to see the highlighted marks on both devices. It could reduce interruption and misorientation during visual sensemaking.

\section{Supporting PC+VR Hybrid Visual Sensemaking}

\begin{figure}
\centering
\includegraphics[width=\columnwidth]{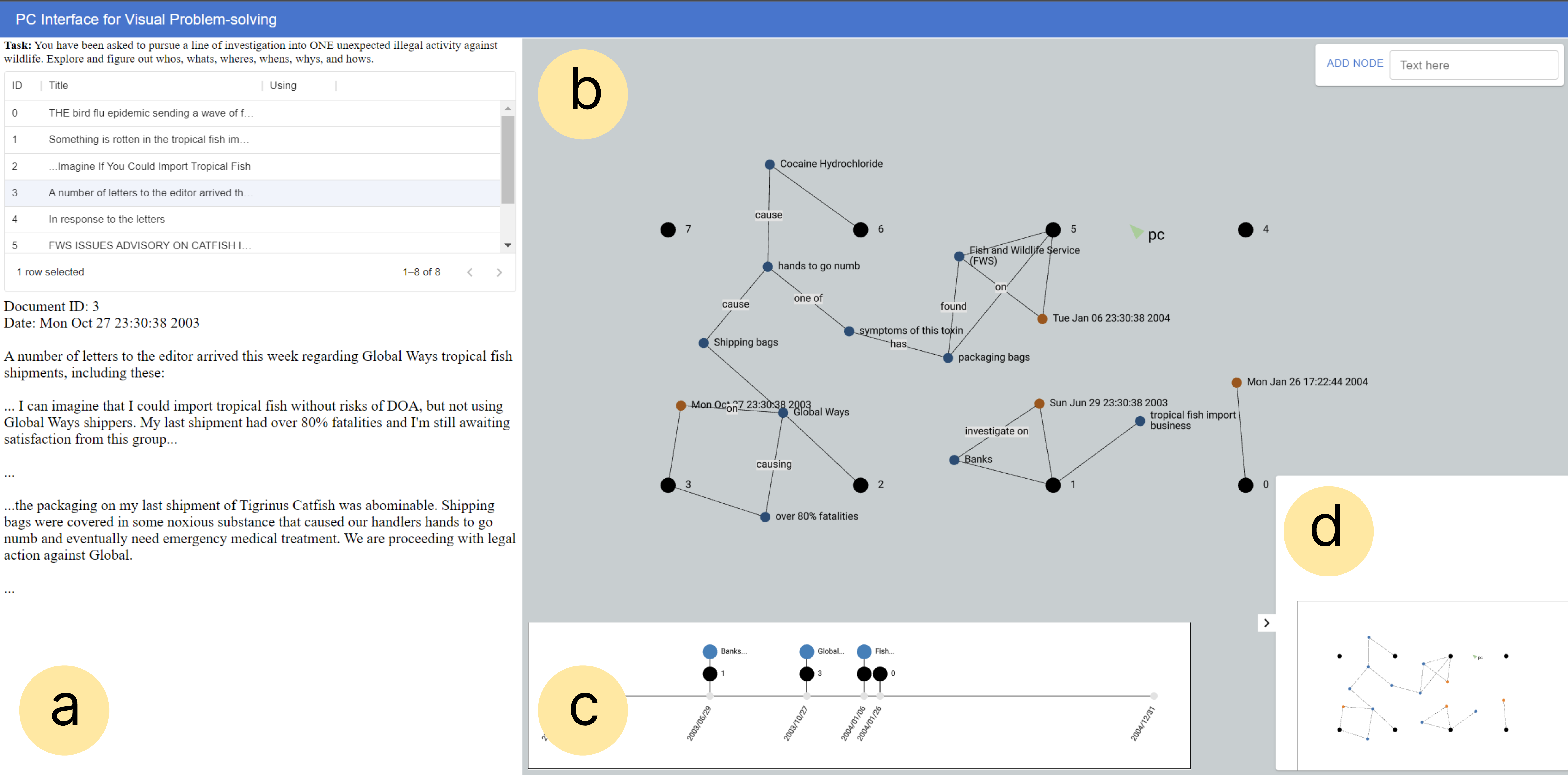}
\caption{Demonstrations of interfaces of PC-only. (a) document view, (b) graph view, (c) timeline view, and (d) minimap view.}
\label{fig:pc_interface}
\end{figure}

We exemplified the design requirements by designing and implementing a hybrid PC+VR system tailored for anticipated visual sensemaking scenarios, specifically crime-solving tasks. 
In this section, we detail our main design considerations and decisions. We begin by introducing the design of PC and VR interfaces, respectively, followed by a discussion of our strategies for integrating them into a seamless interface.

\subsection{PC Interface}
We adapt the designs from previous work~\cite{mahyar2014supporting,tong2023towards} for the PC interface.
The PC interface consists of four views, as shown in \Cref{fig:pc_interface}, document view (a), graph view (b), timeline view (c), and minimap view (d). 
The \textbf{document view} in the PC interface (\Cref{fig:pc_interface}(a)) includes the task description, the document list, and the selected document. The task description provides a clear description of the task that the users are expected to work on and serves as a reminder of the key elements that need to be addressed. The document list displays all available documents, including the document ID and title. Users can select and read a document from the document list. 
The \textbf{visualization view} (\Cref{fig:pc_interface}(b)) serves as a canvas for users to create and read the node-link diagram. In the PC interface, the 2D graph visualizations are displayed. 
Users can add, move, modify, merge, and delete nodes by clicking the mouse and modifying related text using the keyboard in the graph visualization.
Nodes can be added for entities by typing or selecting the relevant text from documents and placing it in the graph visualization. Depending on the text, the node can be classified as a time node and encoded in orange color if the text could be parsed into a date time object without error. Otherwise, the node created will be a normal node and encoded in blue color.
Users can define the relationship between two nodes by adding links, with the links' labels typed or extracted from the documents.
The \textbf{timeline view} (\Cref{fig:pc_interface}(c)) presents time nodes and their connections on a 1D linear timeline.
The timeline is useful for organizing and visualizing time-related information for sensemaking~\cite{mahyar2014supporting}.
The timeline view and the graph are coordinated. Specifically, when users select a node from the graph, the corresponding node in the timeline view is also selected, and vice versa. The view helps users see the nodes in chronological order.
Lastly, a \textbf{minimap} (\Cref{fig:pc_interface}(d)) is presented for users to have an overview of the current graph view. By presenting the graph's current view area, users can better understand their current position and scale relative to the graph.

\begin{figure}
\centering
\includegraphics[width=\columnwidth]{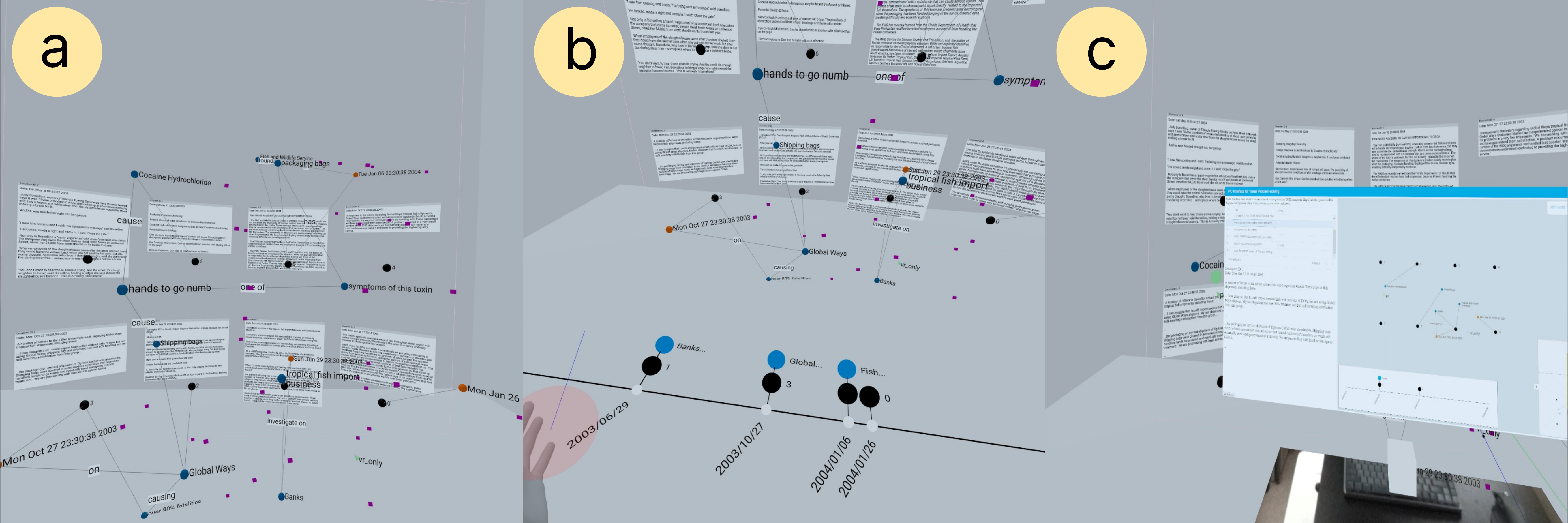}
\caption{Demonstrations of interfaces of (a, b) VR-only and (c) PC+VR (\aka{} hybrid). (a) document view and graph view, (b) timeline view, (c) tracked simulated PC in VR view.}
\label{fig:vr_hybrid_interface}
\end{figure}

\subsection{VR Interface}
To provide the same context in both devices (\hyperref[req:r3]{R3}), we present views with the same functionality and information compared to the PC interfaces, \ie{}, document view (\Cref{fig:vr_hybrid_interface}(a)), graph view (\Cref{fig:vr_hybrid_interface}(a)), and timeline view (\Cref{fig:vr_hybrid_interface}(b)).
At the same time, we adapt the VR user interface designs from a prior work~\cite{tong2023towards} specifically designed for our scenario to optimize the VR interface (\hyperref[req:r2]{R2}).
To utilize the large display space, we distribute the documents in the space in a semi-circular shape, which is commonly adapted by previous work in immersive visualization~\cite{satriadi2020maps,hayatpur2020datahop} and found to be positive to spatial memory~\cite{liu2022effects}, as shown in \Cref{fig:vr_hybrid_interface}(a).

Moreover, the node-link diagram is presented in 3D because 3D node-link diagrams are effective in VR~\cite{belcher2003using,kwon2016study,ware2005reevaluating}.
All graph operations are identical to the PC interfaces.
Furthermore, the text label of nodes and links automatically faces users for their reading. 
To strengthen the spatial relationship between the document and the graph visualization, we externalized the relationship between the created node and the document by providing one black node in front of each document and adding a default link between the created node and the currently selected document nodes (\Cref{fig:vr_hybrid_interface}(a)).

\re{Different from providing the timeline as a 2D panel on the PC, we present the timeline on the floor and utilize the foot interaction for utilizing spatial ability in an immersive environment, as it was found positive for view management~\cite{liu2023datadancing} and used for AR map navigation~\cite{austin2020elicitation}
(\Cref{fig:vr_hybrid_interface}(b)).}{} It allows users to walk on the timeline to select different nodes related to specific moments. Such a design creates an eyes-free interaction and supports secondary navigation tasks so that users might concentrate on the changes in the graph while navigating the timeline.

\begin{figure}
\centering
\includegraphics[width=\columnwidth]{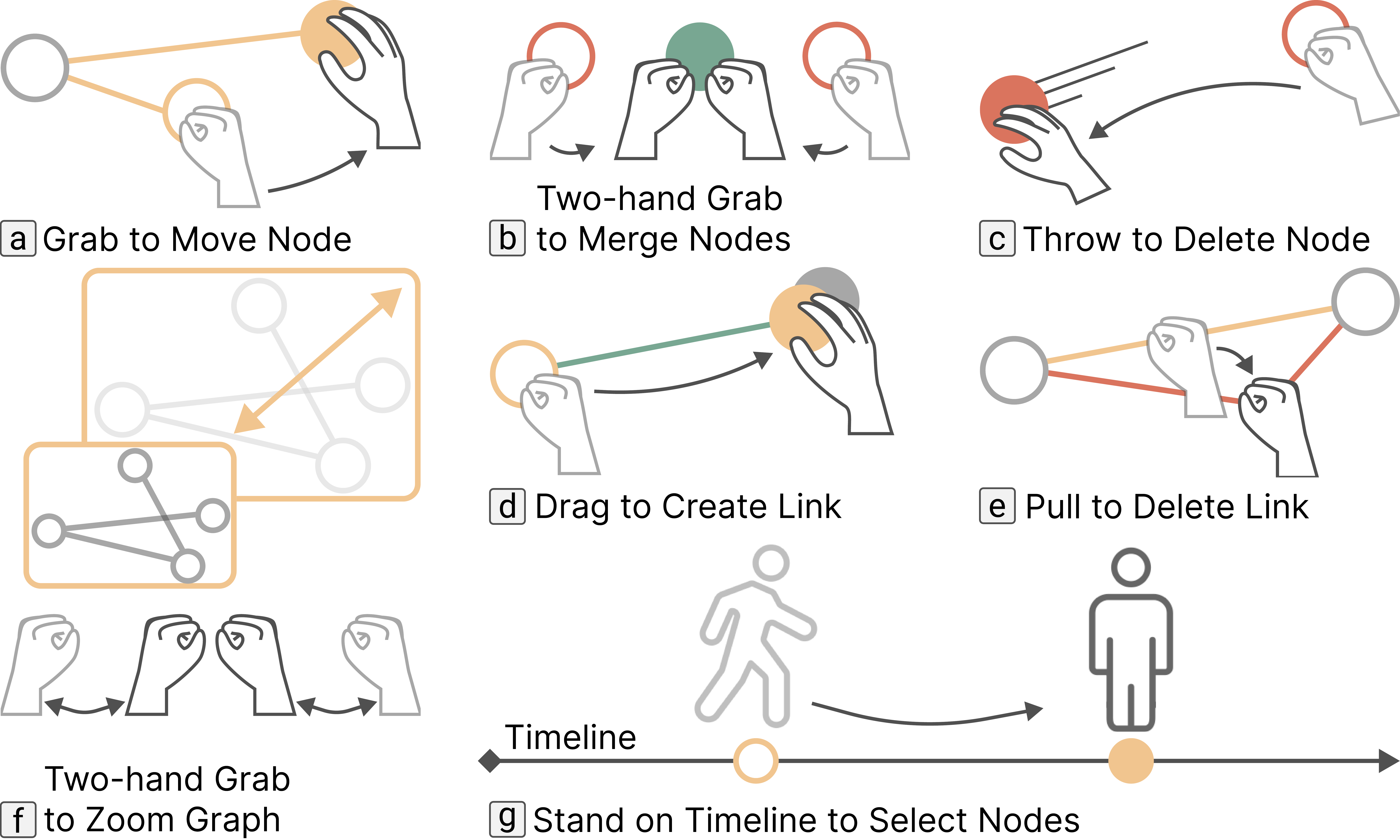}
\caption{The figure shows different hand gestures implemented for graph (a-f) and timeline (g) manipulations. Red nodes indicate that the nodes will be deleted. Green nodes indicate that the nodes will be created. Yellow nodes indicate that the nodes are selected.}
\label{fig:embodied}
\end{figure}

\re{Lastly, we introduce a set of embodied interactions (\ie{}, hand gestures) to nodes and links in VR, as illustrated in \Cref{fig:embodied}, to minimize the context switch between PC and VR devices. It is because the switch between the mouse and keyboard and controllers unavoidably contains three steps: releasing one, locating, and reaching for another, which is less convenient than just releasing the mouse and using hand gestures; in addition, putting the controllers on the table would occupy spaces that interfere with the mouse movement.}{} For the specific designs, a node can be created or updated by selecting the relevant text from the document or using the virtual keyboard text input and placing it in the node-link diagram. Users can ``grab'' the node and move it around (\Cref{fig:embodied}(a)). They can also ``grab'' two nodes and put them together to merge two nodes (\Cref{fig:embodied}(b)) or ``throw'' the node away to delete the node (\Cref{fig:embodied}(c)). Besides nodes, users can define or update any relationship between two nodes by adding a link with labels extracted from the documents or input via a virtual keyboard. Users can ``drag'' one node to another to create an empty link (\Cref{fig:embodied}(d)) and ``pull'' a link to delete the link (\Cref{fig:embodied}(e)). 
To quickly view the graph overview, users can grab the air with both hands and move closer or farther away to zoom in and out on the graph (\Cref{fig:embodied}(f)).
Lastly, users can select a specific node or all nodes with the same date by standing on the corresponding node (\Cref{fig:embodied}(g)) or the white node (\Cref{fig:vr_hybrid_interface}(b)) on the timeline. 

\subsection{Cross-device Interaction in the Hybrid System}
To better support users synchronized using both PC and VR interfaces (\hyperref[req:r4]{R4}) and easy-to-switch input modality and cross-device interaction (\hyperref[req:r5]{R5}), we introduce the following cross-device features and interactions.

\para{Simulated PC in PC+VR.} To reduce the context-switching cost of taking on and off the HMD, we designed a simulated PC, motivated by~\cite{jetter2020vr,immersed2023,hubenschmid2022relive,seraji2024analyzing} (\Cref{fig:vr_hybrid_interface}(c)). 
It allows users to use the PC interface while in the VR environment synchronously (\hyperref[req:r4]{R4}). Users could use a mouse and keyboard to control the PC interface on a physically movable adjustable desk in the VR environment. 
To ensure users could see the keyboard and mouse, we defined a rectangular area below the simulated PC that allows users to see through VR and into reality (\Cref{fig:vr_hybrid_interface}(c) bottom right). 
This see-through area also helps minimize the risk of accidentally bumping into the table and hitting the surrounding area. 
Moreover, to allow users to move the simulated PC in VR, we aligned the position of the simulated PC with the physical desk using a VIVE tracker 3.0, as shown in \Cref{fig:setup}(c). As a result, users could move the simulated PC by moving the desk in reality (\hyperref[req:r1]{R1}).

\para{Synchronized States between Devices.}
To reduce the reload time after switching devices, we synchronize the current document and node selections between interfaces (\hyperref[req:r5]{R5}).
Users can quickly refer back to its current workflow after changing the devices.
For example, users could directly view the document they had last seen in VR when they switched from VR to PC.
Moreover, cross-device linking and brushing are supported. Users could view the same nodes selected from the PC highlighted in VR and vice-versa.
Lastly, we aimed to help users construct a coherent mental model connecting the 2D graph displayed on the simulated PC with the 3D graph in VR.
To achieve this, we prioritized maintaining layout consistency. We initiated this by projecting the 3D graph into a 2D space. Subsequently, we employed a force-directed layout algorithm to minimize the visual clutter.

\para{Hand Gestures as the Main Modality.} To ease the transition between different input interfaces, we decided to use hand gestures (\Cref{fig:embodied}) instead of controllers as the primary interaction modality in the VR interface (\hyperref[req:r5]{R5}). We introduced two main gestures: pinch and grab.
Pinch (air-tap) is the standard gesture for selecting objects using the ray from hand in VR. Grab consists of a fist and a flat gesture for interacting with close-distanced objects. The change from a flat hand to a fist gesture indicates the start of a grabbing action. Conversely, the change from the fist gesture to a flat hand indicates the end of the action. Using hand gestures allows users to easily switch from mouse and keyboard to hand gestures instead of finding and grabbing VR controllers.

\subsection{Implementation}
We used web technology to implement the PC+VR hybrid system,
\ie{}, React.js, d3.js, three.js, and WebXR. To simulate a PC, we performed screen casting from a laptop computer using WebRTC to a plane in VR. Then, to allow users to move the simulated PC in reality, we tracked the desk's movement by placing a VIVE tracker on a wheeled desk and retrieving the pose data using Python OpenVR\footnote{\url{https://github.com/cmbruns/pyopenvr}} and streaming the data to the client web application. 
We set up the initial distance between the physical desk and the tracker for the simulated PC by pressing the ``A'' button of the right-hand VR controller.
To support state synchronization between different devices, we built a server using Node.js and gRPC for fast data communication. 
For gesture recognition beyond the standard pinch gesture, we used Handy.js\footnote{\url{https://stewartsmith.io/handy/}} to recognize fist and flat hand gestures.
Our system code is available at \url{https://github.com/asymcollabvis/hybridvis}.

\section{User Study}
\label{sec:study}

\begin{figure}
\centering
\includegraphics[width=0.6\columnwidth]{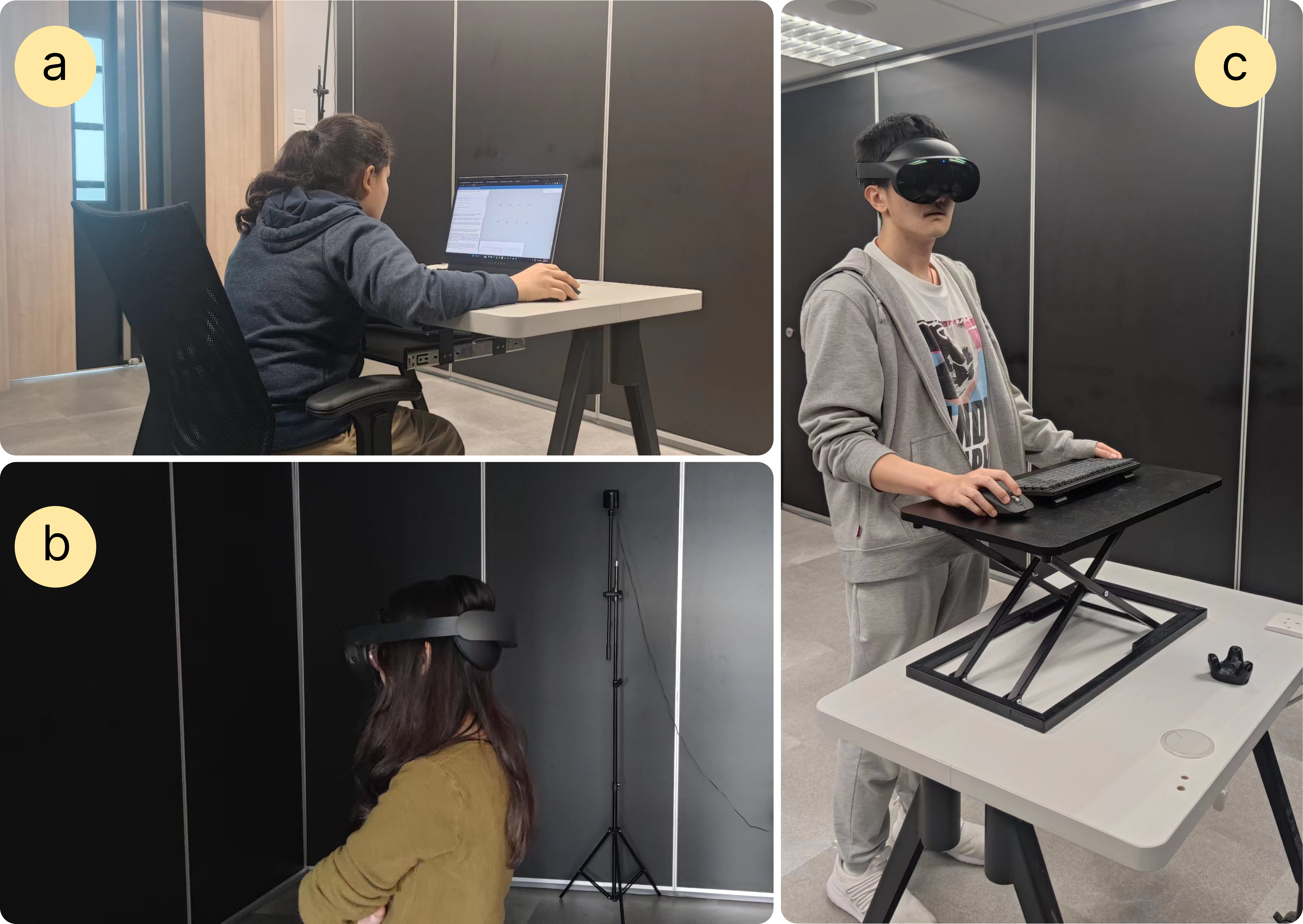}
\caption{The figure shows participants working in PC-only (a), VR-only (b), and PC+VR (hybrid) (c) conditions.}
\label{fig:setup}
\end{figure}

\begin{figure*}
\centering
\includegraphics[width=\linewidth]{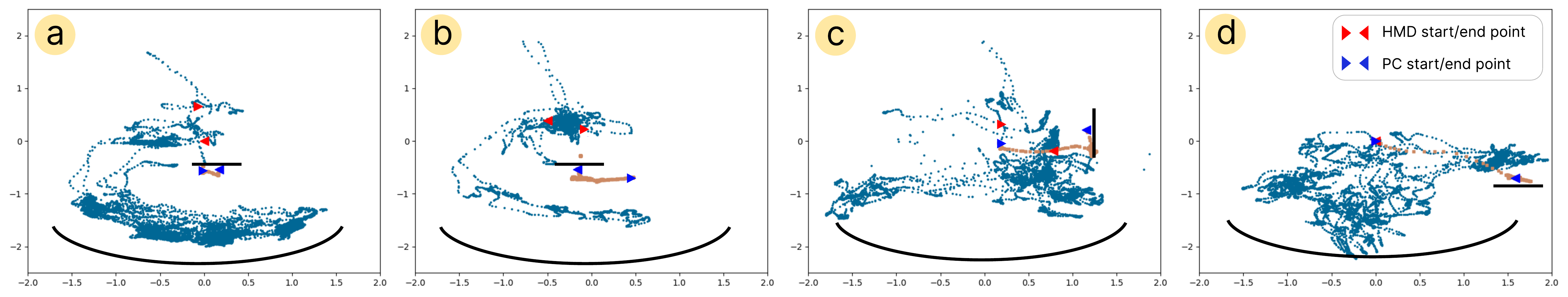}
\caption{This figure shows two representative spatial patterns with an example of corresponding user trajectories (a-d). The black arc represents the position of the documents in VR. The black line indicates the final position and orientation of the PC when they complete the task. 1. Participants did not move the position of the PC and mainly used VR (a) or PC (b) and transitioned to the other when needed. 2. Others moved the position of the PC to the side and used both devices simultaneously with the simulated PC screen perpendicular (c) or parallel (d) to the documents (the black arc).}
\label{fig:patterns_prelimiary}
\end{figure*}

We conducted a\retvcg{}{controlled} user study to 1) \retvcg{investigate}{evaluate} the potential benefits in using a spatial hybrid system (PC+VR), comparing with a single device (PC-only and VR-only), and 2) explore different strategies and device usage patterns in hybrid setups for visual sensemaking. 
The study was approved by the Institutional Review Board \retvcg{(HREP-2024-0111)}{}.

\subsection{Task, Dataset, and Apparatus}
As introduced in \Cref{secc:task}, we tasked our participants with a visual sensemaking puzzle, where the participant needed to create a node-link diagram by extracting entities and relationships from documents to answer a set of questions.  
We used the Blue Iguanodon dataset~\cite{grinstein2007vast} from the VAST 2007 contest, with its difficulty at the graduate level~\cite{whiting2009vast}. 
We used the three subplots with different illegal activities (\ie{}, drug trafficking, wildlife smuggling, and bioterrorism) from previous work~\cite{tong2023towards}. 
To ensure the task was manageable, challenging, and similar to real scenarios, we provided six key documents and added two irrelevant or background documents for each subplot. Finally, we ensured each subplot contained the same amount of documents (\ie{}, eight) with similar total word counts (\ie{}, 1207, 1229, and 1180, respectively).

We used a Meta Quest Pro as the VR HMD and a Dell Alienware x15 R2 Gaming laptop equipped with an Intel i9-12900H CPU, 32GB RAM, an Nvidia GeForce RTX 3070 Ti graphic card, and a 15.6-inch 2560x1440 LCD monitor as the device for both the PC and the backend server.
The study took place in the space of approximately 3$\times$3~meters.

\subsection{Procedure}
The study consisted of four phases and lasted about 120 minutes. The sessions were audio recorded. After completing the whole study, a \$20 Amazon gift card was given as compensation.

\para{Introduction (avg. 5 minutes).} The introduction provided participants with the study's purpose, duration, and setup. We have asked for participants' consent with a consent form before proceeding further.

\para{Main Study and Training (avg. 105 minutes).} The main study phase involved each participant trying three conditions, as shown in \Cref{fig:setup}. 
We controlled the dataset's sequence, and the sequence of the conditions was counterbalanced using the balanced Latin Square method~\cite{bradley1958complete}. Before each condition, we presented a tutorial to participants to help them get familiar with the current condition's interface. Participants were asked to perform all features one by one, including the fact that the table could be moved during the PC+VR condition, and practice altogether to complete a training task.
The training task was designed to help participants understand the study procedure. It was the same as the main study but with a more straightforward dataset of only six documents (439 words in total).
A short interview was conducted to gather feedback on the system's pros and cons for each condition.

\para{Debriefing (avg. 10 minutes).} The debriefing phase involved presenting participants with a questionnaire to rank the conditions in different aspects. Participants' preferences, reasons, usage patterns, and strategies were then discussed in a follow-up semi-structured interview.

\subsection{Pilot Study}
To comprehend the intricacies of our initial design and identify areas that need improvement, we conducted a pilot study before the formal study to explore the initial user experience of our PC+VR system. 
We recruited 12 participants from the local university, including 9 majoring in computer science, 1 in mechanical engineering, 1 in environmental psychology, and 1 in civil engineering. All participants had experience using VR and had previously authored data visualizations before the pilot study.

\para{Key Findings.}
Though we received positive feedback about the PC+VR interface, three key points need improvement.

\vspace{1mm}\noindent\textit{Simulated PC is rarely moved.} With the spatial position logs of the participants and the simulated PC, we identified two patterns: either put the table on the side or stand stationary horizontally and vertically as shown in \Cref{fig:patterns_prelimiary}. The participants did not fully utilize the movement of the table to explore freely in VR. Though participants did not explain it during interviews, we expected that it was due to the heavy weight and large size of the movable table (see \Cref{fig:setup}).

\vspace{1mm}\noindent\textit{Text selection in VR is challenging.} 10/12 participants pointed out that using a long-range ray pointer to select text in VR was not accurate and precise enough. It largely affected the graph creation process during the study. 

\vspace{1mm}\noindent\textit{Readability of text in VR is poor, affecting the usage of simulated PC.} 5/12 participants were not satisfied with the simulated PC in PC+VR due to low resolution for text as well as its interaction latency. 

\begin{figure}
\centering
\includegraphics[width=\columnwidth]{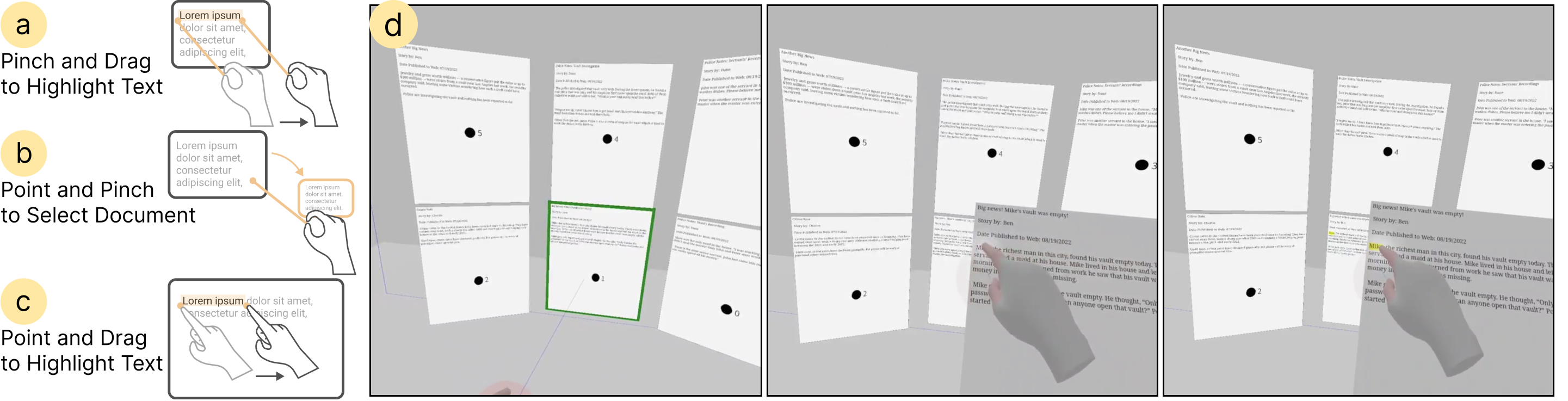}
\caption{This figure demonstrates two techniques for text selection implemented in the VR-only and PC+VR systems: (a) standard text selection with ray pointer and (b-c) two-handed direct touch. (d) demonstrate two-handed direct touch text selection in VR.}
\label{fig:close_interaction}
\end{figure}

\begin{figure}
\centering
\includegraphics[width=0.6\columnwidth]{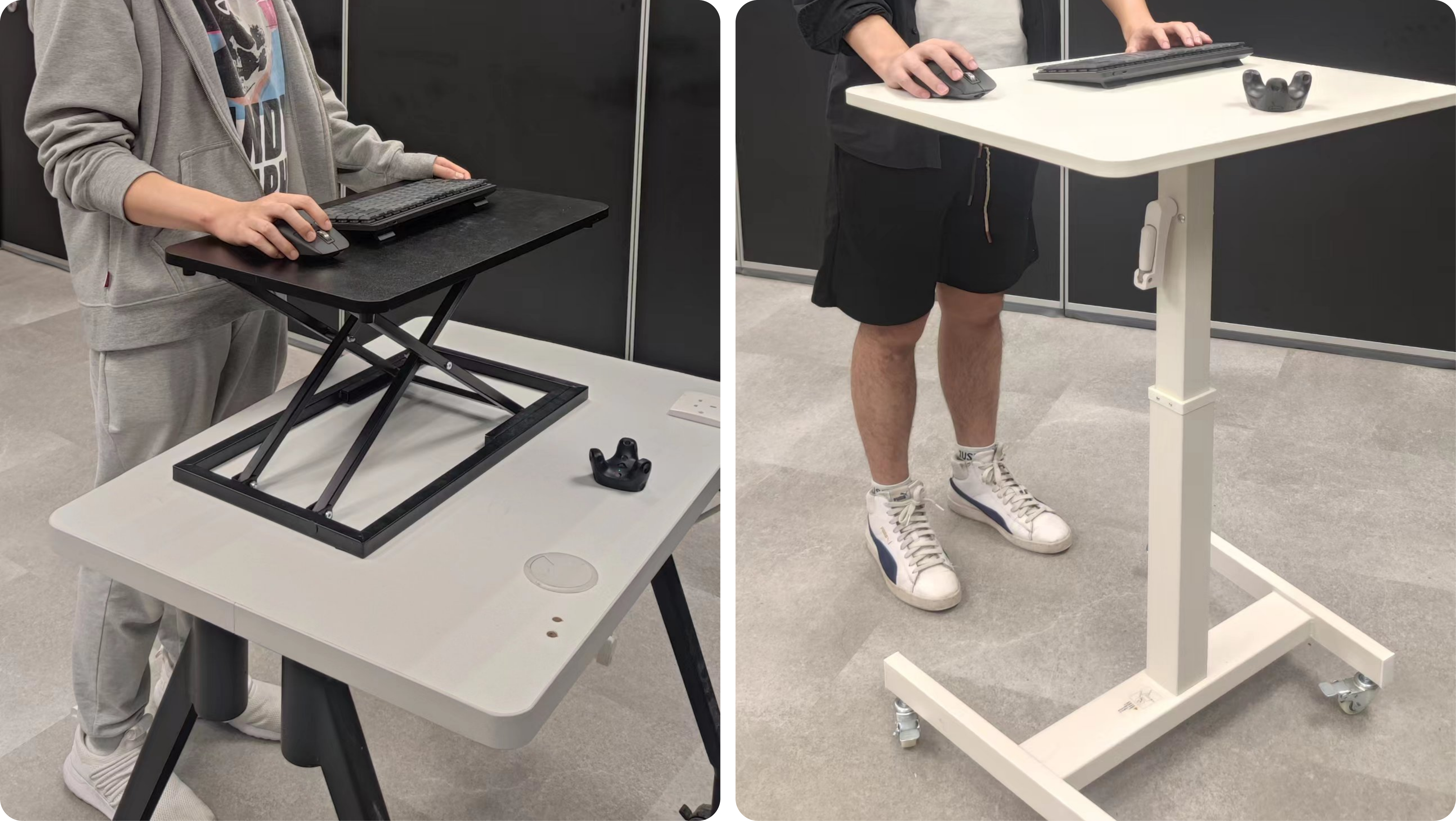}
\caption{Tables used in the pilot (left) and formal (right) studies.}
\label{fig:setup_new}
\end{figure}

\para{Improvements on PC+VR and VR.}
Based on these key findings, we made three corresponding improvements.

\vspace{1mm}\noindent\textit{Using a lighter and smaller movable table.}
To address the transportability issue due to the use of the heavy and large movable table, we decided to use a smaller and lighter wheeled height-adjustable table instead, as illustrated in~\Cref{fig:setup_new}. 
By using a more lighter and smaller table, participants could easily move the table with them in the space, enabling them to switch between environments anytime and anywhere.

\vspace{1mm}\noindent\textit{Introducing close interaction for text selection in VR.}
Performing long-range pointer interactions in VR (\Cref{fig:close_interaction}(a)) can be physically exhausting and suffer from precision, as reflected in our pilot study.
To provide users with an alternative method of selecting the text, we have adapted the two-handed and close-hand interaction approach of Voodoo Dolls~\cite{pierce1999voodoo}. With this approach, users can pinch to select a document using their left hand (\Cref{fig:close_interaction}(b)). 
A smaller version of the document view will then be duplicated and mirrored onto their hand, simulating the experience of picking up a piece of paper. Users can then use their right-hand index fingertips to touch and drag over the text they want to select on the duplicated document (\Cref{fig:close_interaction}(c)). 
We also added a light source to the index fingertip to provide a visual cue for enhanced depth perception (\Cref{fig:close_interaction}(d)).

\vspace{1mm}\noindent\textit{Improving the rendering of text and simulated PC display in VR.} 
To ensure sharper text and a clearer simulated PC screen display, we applied WebXR layers\footnote{\url{https://www.w3.org/TR/webxrlayers-1/}} to render the text and the real-time display of the simulated PC for both the VR and PC+VR prototypes.
Using WebXR layers can improve performance by significantly reducing the rendering rate of static text, as well as increasing the image quality by direct rendering to the final buffer without double sampling and distortions. 
The original simulated PC was sized around 85'' with 4K resolution to accommodate text legibility in the pilot study, and participants complained that it was bulky and created unnecessary occlusions. Thanks to the enhanced graphic rendering, we were able to provide a smaller simulated PC screen, now at 32'' with 1440p resolution \retvcg{(0.045 visual angle per pixel\footnote{\retvcg{Calculated with 1m eye-to-screen distance. The visual angle per pixel of the actual 32'' 1440p monitor is 0.015. The lower the better. The detailed calculations can be found in the supplementary material.}{}})}{}. This enhancement reduces the overlap between the simulated PC and the VR content and minimizes interaction latency while ensuring text clarity.

\subsection{Participants}

\re{In terms of the participants in the main user study, we tried to strike a balance between the participants' diversity and the required expertise. We recruited users with different backgrounds and excluded those who had no experience in data visualization or VR to reduce the learning effect.}{} Finally, we recruited 18 participants
 who had not joined the pilot user study.
There were ten males and eight females, with a mean age of 25.6 (SD = 3.73). All participants were graduate students from different majors: computer science (12), linguistics (1), material (1), art (1), bioinformatics (1), construction management (1), and physics (1). 
\re{The distribution of experience in using VR was within one year (12), 1-2 years (2), and more than two years (4). 
The distribution of experience in using data visualization was within one year (8), 1-2 years (5), and more than two years (5).}{}

\subsection{\retvcg{Exploratory Hypotheses}{Hypotheses}}
\retvcg{Our overarching goal is to explore the differences in user behaviors and experiences between using a hybrid system and a single environment. To systematically guide our exploration, we developed several representative hypotheses based on previous empirical results, our pilot study, and the design of our conditions (see \Cref{sec:pc_vr_design}). These exploratory hypotheses serve as a structured starting point for focused observation and discussion, rather than for performance comparisons.}{}

\para{Accuracy (\textit{H\textsubscript{acc}}).} \label{hypo:acc} We did not expect any difference in accuracy as the same functionalities were consistently provided in all testing conditions.

\para{Time (\textit{H\textsubscript{tim}}).} \label{hypo:tim} We expected PC-only to outperform VR-only and PC+VR based on previous studies, which found desktop interactions faster than VR interactions due to less required movement~\cite{bach2017hologram,chen2012effects,wagner2018immersive,arms1999benefits}. Compared to VR-only, PC+VR can partially benefit from faster desktop interactions, especially for precise interactions like selecting the text, but the extra context-switching may introduce more time costs.

\para{User Experience (\textit{H\textsubscript{exp}}).} \label{hypo:exp} We believed PC+VR could have less task load than PC-only and VR-only, as participants had the choice to choose the optimal device for the given task components.
However, PC+VR may introduce distraction when switching between devices, influencing concentration.

\para{Number of Interactions (\textit{H\textsubscript{int}}).} \label{hypo:int} Our task is interaction-intensive, which requires foraging and structuring information. We anticipated VR-only and PC+VR would require fewer interactions than PC-only, based on previous investigations~\cite{lisle2020evaluating,in2023table}, possibly due to the unlimited display space in VR requiring fewer navigations. 
Between VR-only and PC+VR, the advantage of performing precise interactions using the PC in PC+VR can lower the required number of interactions.

\para{User Preference (\textit{H\textsubscript{pre}}).} \label{hypo:pre} We considered that participants would mostly prefer PC+VR over PC-only and VR-only due to the existing limitations of a single computing environment.

\subsection{Measures}
We collected data from the formal study with the enhanced version of the spatial hybrid system.
We recorded the \textit{time} taken to complete each task, task \textit{accuracy}, and the \textit{number of interactions} performed to complete the task.
In terms of interactions, we considered all graph-related interactions, including adding, removing, and updating nodes/links, as well as merging nodes.
We used the \textit{NASA TLX task load}~\cite{sandra2006nasatlx} and adapted \textit{concentration}~\cite{novak2000measuring} questionnaires to collect subjective ratings of participants' user experience.
We also asked participants to \textit{rank} the three conditions based on their preference for four different task components: \textit{authoring}, \textit{exploring}, \textit{discovering}, and \textit{interaction}, as well as their \textit{overall} preference at the end of the study. 
Qualitative feedback from the debriefing interviews was used as evidence to explain task efficiency, ratings on user experience, and preference.
We logged users' spatial movements in the space to contribute insights and empirical understanding about how people use PC+VR hybrid interfaces.
Moreover, we also tracked the HMD’s head gaze data to detect the objects users are currently looking at in VR.

\section{Results}
\label{sec:result}

\begin{figure*}
\centering
\includegraphics[width=\textwidth]{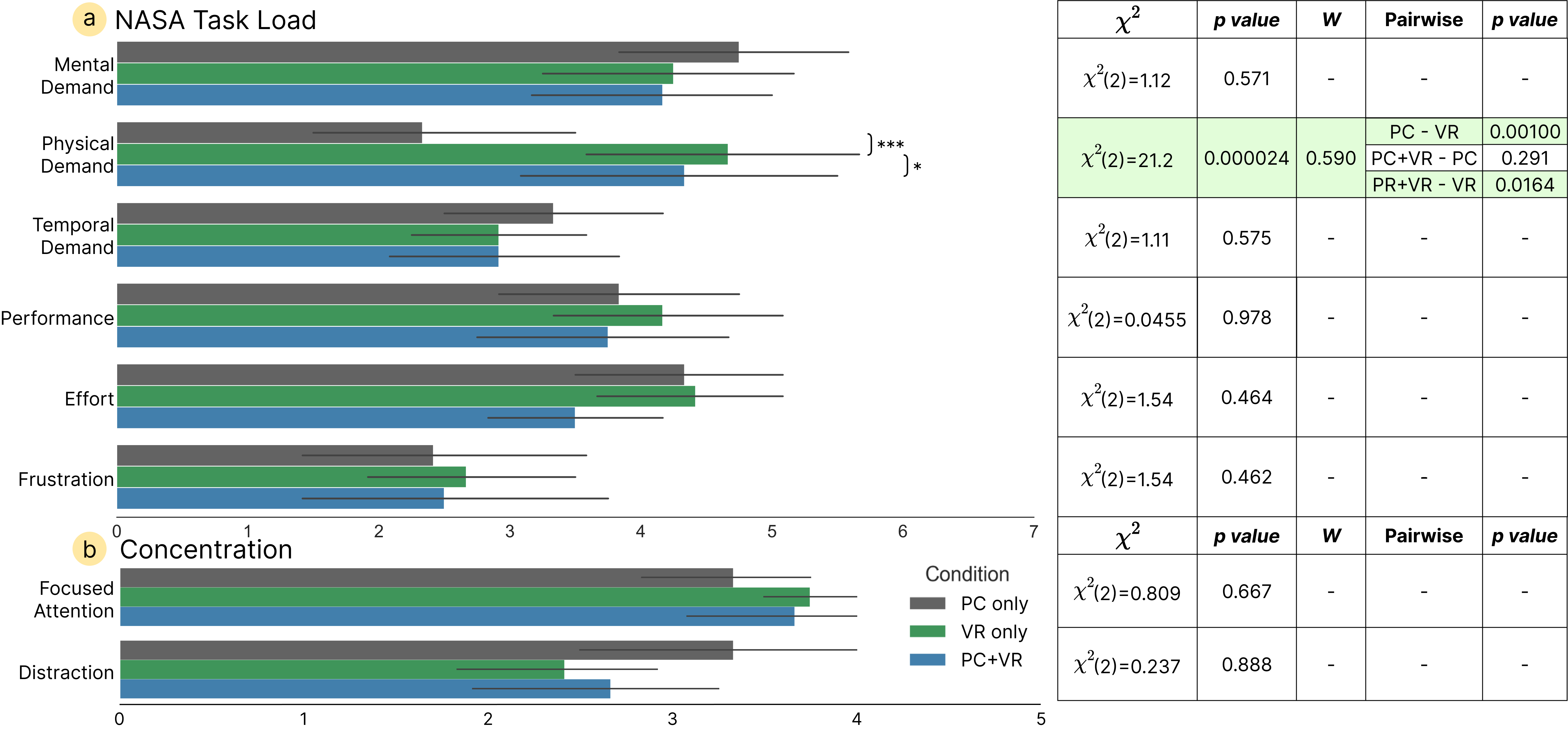}
\caption{Results of the NASA Task Load questionnaire (a) and two questions for users' concentration (b). Error bars show the 95\% confidence interval (CI).
Significance values are reported as $p {<} .05(*)$ and $p {<} .001({*}{*}*)$. \re{The table presents the statistical data. Significance values ($p {<}.05$) are highlighted in green.}{}}
\label{fig:user_experience}
\end{figure*}

\begin{figure*}
\centering
\includegraphics[width=\textwidth]{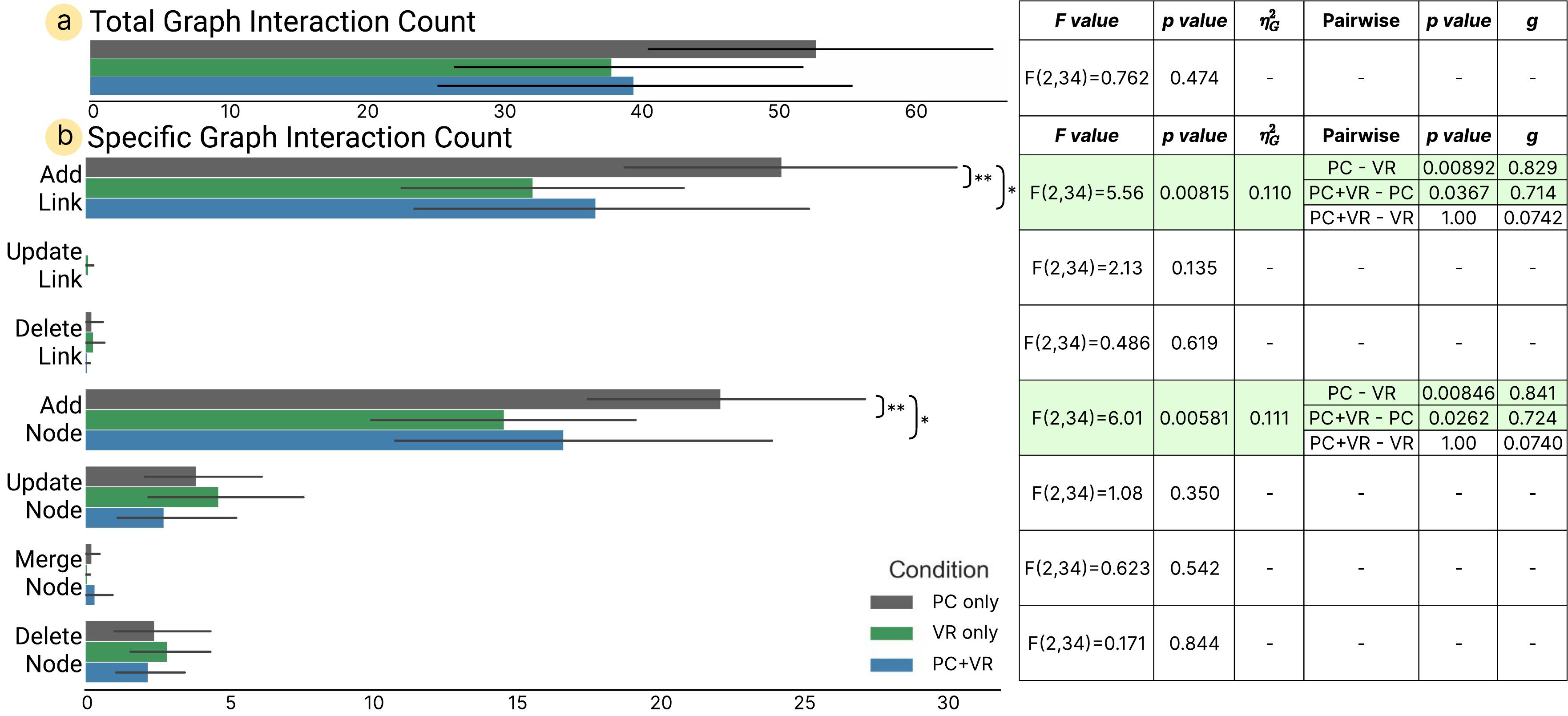}
\caption{The average number of interactions logged for each condition. Error bars show the 95\% confidence interval (CI). 
Significance values are reported as $p {<} .05(*)$ and $p {<} .01(**)$. \re{The table presents the statistical data. Significance values ($p {<}.05$) are highlighted in green.}{}}
\label{fig:interaction_result}
\end{figure*}

For \textit{time} and \textit{number of interactions}, we first applied a log transformation to check the normality assumption. We then employed \textit{repeated measure ANOVA} to evaluate the effect of the three study conditions on the dependent variables. 
We determined the significance of including an independent variable or interaction terms using a log-likelihood ratio.
Additionally, we performed post-hoc pairwise comparisons using \textit{t-test with Bonferroni correction}.
For other non-parametric data, such as accuracy, user experience ratings, and rankings, we conducted \textit{Friedman tests} with \textit{Nemenyi post-hoc} analysis.
For qualitative feedback, we used affinity diagramming~\cite{hartson2012ux} to organize and analyze the transcripted recordings.

\para{Time}, \textbf{Accuracy} and \textbf{User Experience Ratings}.
With repeated measure ANOVA, we did not find significance for \textit{time} ($F(2,34){=}1.53, p{=}0.230, \eta^2_G{=}0.0537$).
With the Friedman test, we also did not find significance for \textit{accuracy} ($\chi^2(2){=}0.426,p{=}0.808,W{=}0.0118$) and \textit{user experience ratings}, \ie, concentration and NASA Task Load except physical demand, as shown in \Cref{fig:user_experience}. VR ($avg{=}4.33, CI{=}0.950$) is significantly more physically demanding than PC+VR ($avg{=}2.83, CI{=}0.857, p{=}0.0164$) and PC ($avg{=}1.67, CI{=}0.418, p{=}0.00100$).

\para{Number of Interactions.}
We did not find that our testing conditions significantly affected the total number of interactions, \Cref{fig:interaction_result}(a).
However, when we subdivided the number of interactions into specific types, see \Cref{fig:interaction_result}(b), we did find significant differences in adding nodes ($F(2,34){=}6.01, p{=}0.00581, \eta^2_G{=}0.111$) and links ($F(2,34){=}5.56, p{=}0.00815, \eta^2_G{=}0.110$) reflected.
These two interactions were the primary interactions performed by the participants to build the graphs.
By testing the pairwise significance, we found that participants added more nodes in PC-only ($avg{=}22,CI{=}4.76$) than VR-only ($avg{=}14.5,CI{=}4.83,p{=}0.00846$) and PC+VR ($avg{=}16.6,CI{=}6.63,p{=}0.0262$), as well as more links in PC-only ($avg{=}24.1, CI{=}5.89$) than VR-only ($avg{=}15.5, CI{=}5.26, p{=}0.00892$) and PC+VR ($avg{=}17.7,CI{=}6.97,p{=}0.0367$).

\begin{figure*}
\centering
\includegraphics[width=\textwidth]{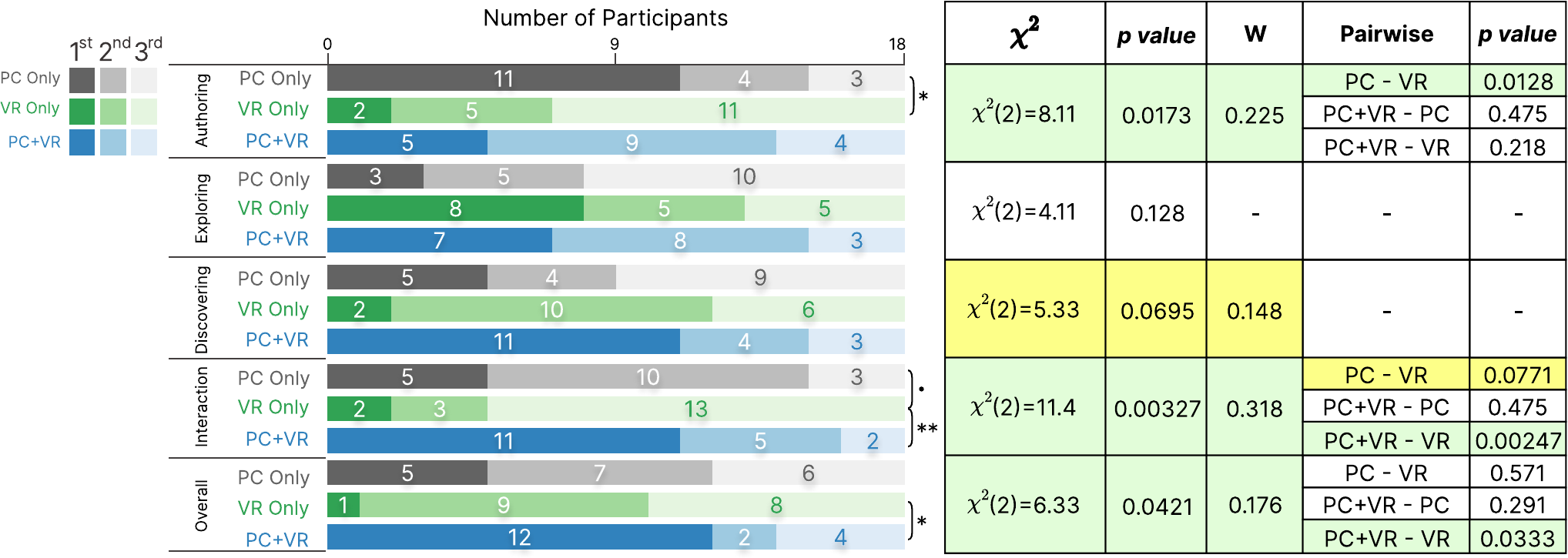}
\caption{Ranking results in terms of Authoring, Exploring (finding a target), Discovering (finding an insight), Interaction, and Overall ranking of the post-study survey for the three conditions in the formal study.
Significance values are reported as $p {<} .05(*)$ and $p {<} .01(**)$.
Marginal significance values are reported as $.05{<}p{<}.1(\cdot)$
\re{The table presents the statistical data. Significance values ($p{<}.05$) and marginal significance values ($.05{<}p{<}.1$) are highlighted in green and yellow, respectively.}{}}
\label{fig:ranking}
\end{figure*}

\para{User Preference.} 
As shown in \Cref{fig:ranking}, there are significant differences in terms of overall preference ($\chi^2(2){=}6.33,p{=}0.0421,W{=}0.176$), interaction ($\chi^2(2){=}11.4,p{=}0.0033,W{=}0.318$), and authoring experiences ($\chi^2(2){=}8.11,p{=}0.0173,W{=}0.225$). \retvcg{Overall, participants liked PC+VR more than VR-only (12/18,$p{=}0.0333$). In particular, in terms of interaction, PC+VR is ranked higher than VR-only (11/18,$p{=}0.00247$).}{} We also found that PC-only is preferred in interaction slightly more than VR-only (5/18,$p{=}0.0771$).
\re{Moreover, more participants preferred authoring with PC-only than VR-only (11/18, $p{=}0.0128$).
Though we did not observe further significant differences, there is a marginal difference in terms of discovering ($\chi^2(2){=}5.33,p{=}0.0694$), and more participants preferred discovering with PC+VR (11/18).}{}

\para{Qualitative Comments and Different Strategies Used.}
Participants raised different opinions towards different interfaces.
PC interface was described as \userquote{easy and simple to interact with the graph precisely} (16/18), VR interface was \userquote{suitable for reading documents} (15/18), and PC+VR could \userquote{leverage benefits from both of them and overcome their weaknesses} (12/18).
With the improvement and involvement of more participants than in the preliminary study, we received new points for our conditions. 
Though with a higher resolution of text, more participants (from 25\% to 44\%) considered the PC interface unsuitable for reading documents (8/18) for two major reasons. The first one is that the document list is inefficient in switching between documents (5/8), and the small screen of the simulated PC interface is unsuitable for reading (4/8). For example, P3 commented that \userquote{I need to click and jump to the different articles to find the association in PC.} P16 added that \userquote{the PC screen size is too small [compared to VR] which is not good for reading.} For the PC+VR condition, we did not receive complaints about our simulated PC regarding its latency and low resolution.

\re{Moreover, on the one hand, 13 participants mentioned that they had used a different strategy in using different interfaces for completing the task. Specifically, ten participants mentioned that they utilized graphs more in PC-only, while in VR-only and PC+VR mode, they tended to read all documents first and create the graph only with key documents. For example, P3 mentioned that \userquote{I tend to create [graphs] on the PC. In VR, I only need to remember spatial positions and don't want to create graphs.} P18 added that \userquote{The operation in PC is very familiar and precise, and I tend to record more information (using the graph) because the interaction cost is relatively small, and I hope I won’t have to go back and switch documents. In VR, I didn't create graphs first; I read all the documents first and then only created graphs for key information.} On the other hand, four participants mentioned that the strategy used in all conditions is similar. For example, P7 stated that she analyzed the time first, then read documents [based on the time], and created the graph for all three conditions. Moreover, P6 mentioned that the strategy used was independent of the interface but experience of using visual sensemaking tools.}{}

\para{PC+VR Hybrid User Strategies.}
\re{To analyze the user strategies of the PC+VR interfaces, we visualize the interaction logs and}{We} group participants into different strategy categories from the perspectives of the \emph{temporal} and \emph{spatial}.

\begin{figure}
\centering
\includegraphics[width=\columnwidth]{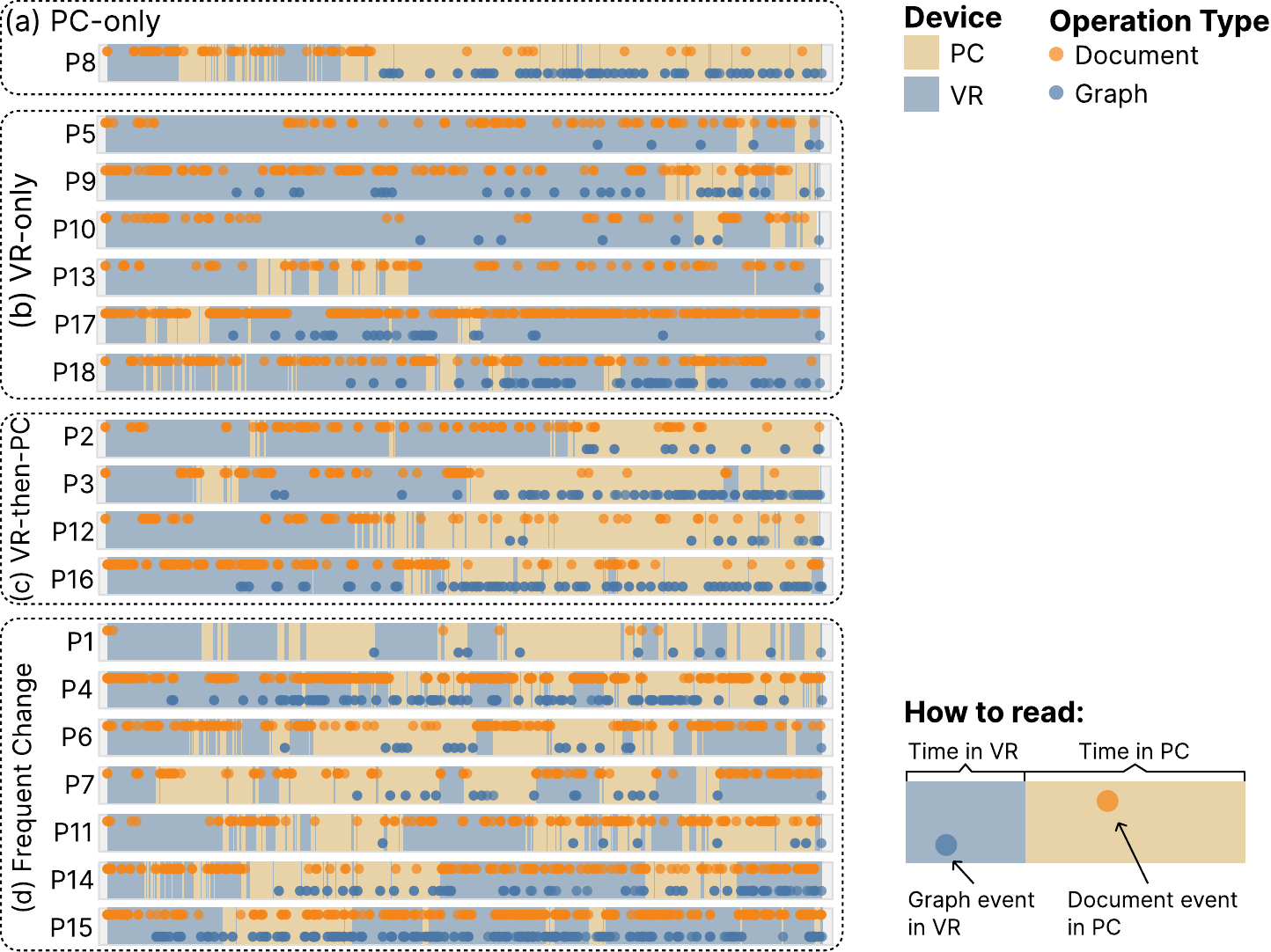}
\caption{The distribution of time for all participants in using PC and VR in the PC+VR condition in the formal study: primarily use PC (time of PC ${>} 75\%$) (a), primarily use simulated VR (time of VR ${>} 75\%$) (b), used both VR and simulated PC with non-frequent switching (c), and used both VR and simulated PC with frequent switching (d). The orange and grey dots represent the events related to documents and graphs.}
\label{fig:time}
\end{figure}

\begin{figure*}
\centering
\includegraphics[width=0.8\linewidth]{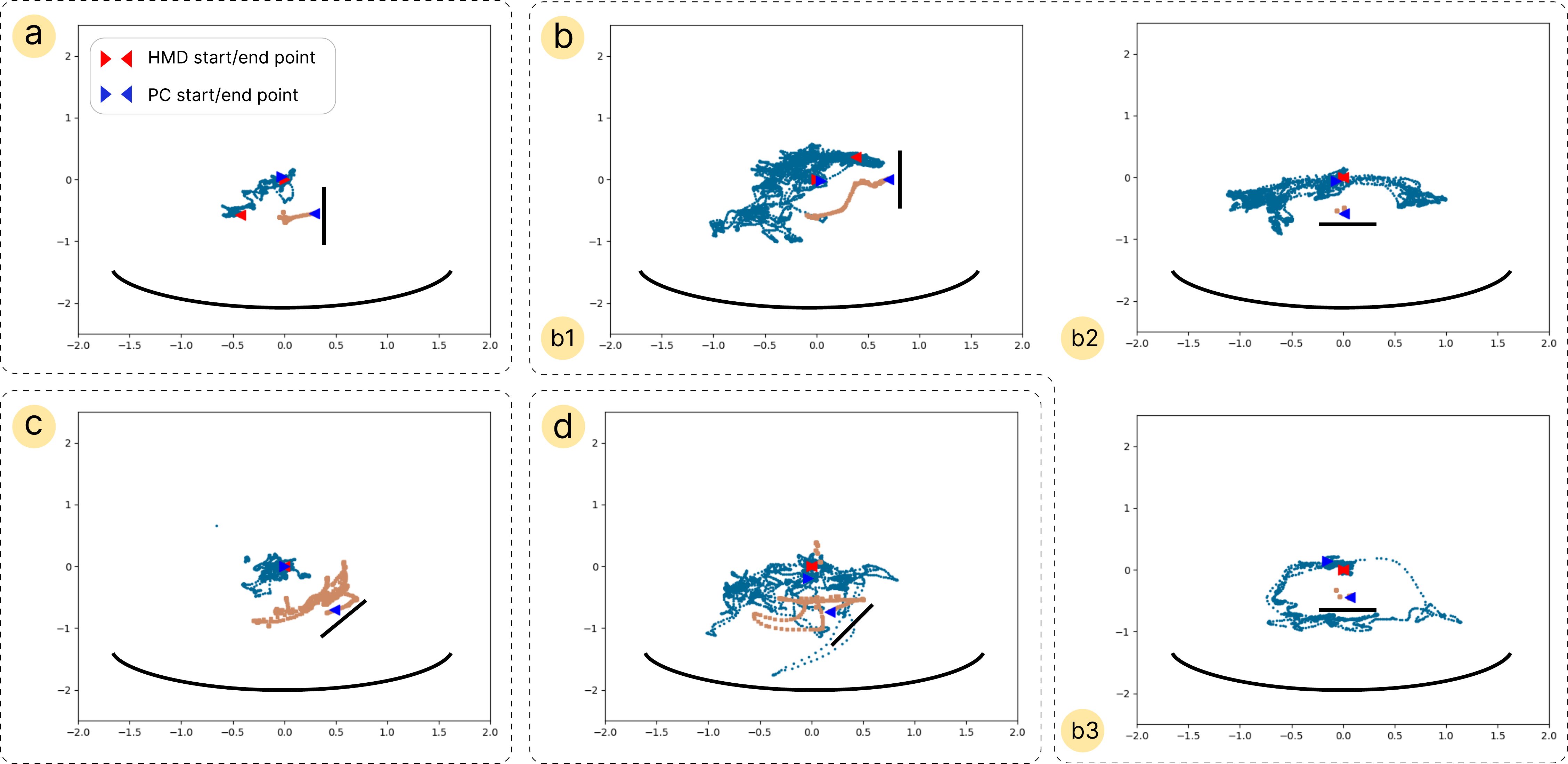}
\caption{This figure shows four representative patterns of how users solved the task with an example of corresponding user trajectories (a-d) in the formal study. The black arc represents the position of the documents in VR. Some participants did not move the position of the PC and themselves (a) or just themselves (b). Others moved the position of the PC but remained stationary (c) or together (d).}
\label{fig:patterns}
\end{figure*}

\vspace{1mm}\noindent\textit{Temporal strategies.}
We were interested in how much time the participants spent in the different environments and how frequently they switched between environments. 
We identified four different strategies from our participants based on 1) the time spent on PC or VR and 2) the frequency of switching between devices:
\begin{enumerate}[noitemsep,topsep=0pt,parsep=0pt,partopsep=0pt,leftmargin=*]
    \item \textbf{PC-only}: 1/18 participant spent most of their time (${>} 75\%$) in the PC environment, \Cref{fig:time}(a).
    \item \textbf{VR-only}: 6/18 participants spent most of their time (${>} 75\%$) in the VR environment, \Cref{fig:time}(b).
    \item \textbf{VR-then-PC}: 4/18 participants spent noticeable time in each environment without frequently switching, especially having a pattern of VR first, then PC, \Cref{fig:time}(c).
    \item \textbf{Frequent Switch}: 7/18 participants spent similar time in each environment and frequently switched environments, \Cref{fig:time}(d).
\end{enumerate}

\vspace{1mm}\noindent\textit{Spatial strategies.}
We also wanted to know how participants moved the PC as well as how they moved in space.
We also identified four different strategies from our 18 participants based on 1) the movement of the user (${\geq} 290m$ or ${<} 290m$) and 2) the movement of the table (${\geq} 5.5m$ or ${<} 5.5m$):
\begin{enumerate}[noitemsep,topsep=0pt,parsep=0pt,partopsep=0pt,leftmargin=*]
    \item \textbf{Stationary User and PC}: 3/18 participants almost did not move the PC and primarily stood near the initial starting position, \Cref{fig:patterns}(a).
    \item \textbf{Stationary PC}: 8/18 participants almost did not move the PC but moved themselves to use the space in VR, \Cref{fig:patterns}(b). With these eight participants, we also observed three subpatterns: \textit{\underline{PC Side}} (3/8, move the PC to the side, \Cref{fig:patterns}(b1)),  \textit{\underline{User Side}} (3/8, move towards right or left sides, \Cref{fig:patterns}(b2)), and \textit{\underline{Circle}} (2/8, move around the PC, \Cref{fig:patterns}(b3)).
    \item \textbf{Self-Rotation}: 5/18 participants remained stationary while they moved the PC in the space, \Cref{fig:patterns}(c).
    \item \textbf{Carrying}: 2/18 participants constantly moved the PC with them in the VR space, \Cref{fig:patterns}(d).
\end{enumerate}

\section{Discussions}
\label{sec:discussion}

\para{Hybrid PC+VR interface was preferred.}
After improving the usability of the simulated PC, we observed a significant preference towards hybrid systems overall, particularly interaction. \textit{H\textsubscript{pre}} is supported.
\re{Most participants preferred the PC+VR hybrid interface, as it effectively combined the strengths of both devices, especially
when completing a complicated sensemaking task that requires both an overview and a detailed view of the information for navigation, foraging, insight generation, or synthesizing. 
VR provides participants with an overview of documents so they can quickly scan and move in space to read the documents and graphs with preferred locations and angles.
At the same time, PC offers a detailed view by supporting precise interactions and a compact view for digging deep into specific documents for graph construction.}{}

\re{\para{Hybrid PC+VR help relieving physical demand.}
While the PC+VR setup did not yield a marked improvement in the overall user experience---as evidenced by ratings on mental load and effort are similar for all conditions, 
except for physical demand (thus, \textit{H\textsubscript{exp}} is not supported)---it is noteworthy that the hybrid interface exhibited lower physical demand than a purely VR setup. Ratings on mental load and effort are similar for all conditions, indicating that the hybrid systems do not introduce new mental loads and efforts.}{}
This is worth noting, given that participants in both scenarios operated within a VR environment requiring spatial movement for navigation. 
\re{A possible explanation is that the predominant physical demands in VR stem from interactions, and the smooth transition between hand gestures and mouse and keyboard promotes streamlined interactions, particularly for those that require high precision. Moreover, the wheeled table surface could reduce arm movement, which increases comfort and reduces physical strain~\cite{cheng2022comfortable}. Future designs could consider involving a wheeled desk to support spatial navigation while reducing physical demands for hybrid systems, as well as VR systems.}{}

\para{Hybrid PC+VR interface with spatial features did not hinder performance.}
The study found no significant differences in time, accuracy, and concentration across conditions. Compared with the previous similar result~\cite{pavanatto2021we}, we further found that the involvement of spatial navigation \retvcg{with the movable simulated PC}{} in VR did not hinder performance when using the hybrid interface.
In particular, the task accuracy remains the same as we expected since the functionalities in all conditions were the same (\textit{H\textsubscript{acc}} is supported).
Initially, we had anticipated that PC would outperform VR and PC+VR in terms of speed, but the results suggest that both PC+VR and VR-only performed the same as PC-only. \textit{H\textsubscript{tim}} is largely inconclusive. 
This could be attributed to its intuitive embodied interactions, which enabled participants to physically navigate and leverage their spatial memory to improve their performance. 
Furthermore, the VR interface's ability to display all documents in a layout that allows physical navigation may have contributed to its better-than-expected performance.
More specifically, participants could just rotate their heads to scan, read, and search documents.
Such benefits of physical navigation have been verified in the context of large display walls~\cite{ball2007move}.
Moreover, user concentration did not significantly differ from hybrid PC+VR to single PC or VR conditions. We found no evidence for \textit{H\textsubscript{con}}. Combining the result that PC+VR was observed to have similar time performance, it indirectly demonstrates the success of our efforts to minimize context-switching costs between PC and VR.
\re{Future hybrid systems for data visualization could consider involving more spatial ability since we found evidence that it increased user preferences while not hindering performance.}{}

\para{PC+VR and VR-only required fewer number of \re{nodes and links}{interactions} to complete the study task than PC-only.}
Though we did not find a significant difference in the total number of interactions between conditions, we found a significant difference for the two main interactions, \ie{}, adding nodes and links, between VR-only and PC-only, and between PC+VR and PC-only.. (\textit{H\textsubscript{int}} is partially supported).
Two reasons could explain this finding. 
First, participants tended to add more nodes and links to store information from documents to reduce navigation in PC-only.
\re{As previous studies suggested, the unlimited display space in VR helps reduce navigation, while the limited display space in PC could require more navigation. Since only one document can be viewed at a time with limited display space on a PC, participants need to create more nodes and links to store the information using the graph to reduce the number of navigation between documents.}{} In contrast, documents were displayed with details and spatially distributed in VR, seamlessly blending into the room-sized visualization. This allowed participants to create fewer visualization elements to externalize document information. It could potentially indicate that PC+VR and VR-only facilitate a more precise graph for a more complex task.
Second, authoring the graph in VR is more difficult than it is on PCs and is unfamiliar to users. Although we provided an alternative text selection technique using close interaction, it is still difficult for users to select text precisely with mid-air gesturing due to hand shaking. Therefore, participants might try to create a more precise graph with fewer interactions in the VR environment.
More studies could investigate larger graphs or more complex problems to evaluate these possible reasons.

\para{Temporal strategies.} Most participants (11/18) used both environments in a complementary way. We observed this in two levels of granularity: interaction and task levels.
\textit{Interaction levels}: 
For a given low-level interaction, participants (7/11) would choose the most suitable environment. For example, some participants found VR helpful for exploring the node-link diagram and walking around to find insights but switched to PC to create the link because text selection and input were easier on PC. 
Participants who adopted this strategy switched between environments more frequently, as demonstrated in \Cref{fig:time}(d). \textit{Task levels}: Task-level complementation is more strategic, which might be the best fit for previous transitional approaches~\cite{jansen2023autovis,hubenschmid2022relive}. Participants (4/11) planned ahead and chose the best environment for different sensemaking stages.
For instance, participants first read documents in VR to get a general overview of the story, and then they extract keywords to validate their thoughts using the PC. 
These participants only switched between environments a few numbers of times, as demonstrated in \Cref{fig:time}(c).
Adopting such a strategy might seem counter-intuitive, given that smaller displays are typically associated with overviews, while larger ones are often linked to detailed views. 
However, participants seem to harness the spatial awareness and memory offered by VR to maintain a superior mental model of spatial information, leading to enhanced wayfinding performance---essential attributes of an overview. Concurrently, the precision and inspection capabilities of the PC make it well-suited as a detailed view.
The findings resonate with Schneiderman's mantra of information seeking ``\textit{Overview First, Zoom and Filter, Then Details-on-Demand}''~\cite{shneiderman1996eyes} and the design goals of the hybrid interface to combine immersive display and high-resolution input capabilities~\cite{feiner1991hybrid}.
\retvcg{While some participants transitioned only a few times at the task level or even remained in a single environment (11/18), a substantial portion of participants (7/18) frequently switched between devices at the interaction level. This highlights the need to reduce the transitional costs between devices. Future designs should consider non-transitional approaches~\cite{immersed2023, wang2020towards, pavanatto2021we, wang2022understanding} to lower these costs.}{}

\para{Spatial strategies.} With equipment of a lighter and smaller table, it opens more possibilities for spatial movement for a hybrid user interface. Interestingly, the spatial patterns reflect the current usage of computing devices. \textit{Stationary User and PC} mimicking the stationary working environment of sitting and watching multiple monitors, when \textit{stationary PC} is the representation of the current workflow of the hybrid system, such as ReLive~\cite{hubenschmid2022relive}. Users work in the space in VR and go to a specific location to work on the PC. Besides, two additional patterns, \textit{self-rotation} and \textit{carrying}, were observed compared to the preliminary study. \textit{Self-rotation} captures participants who rotate the desk and the PC screen while using the spatial distribution of documents in VR, while \textit{carrying} describes participants who fully unleash the potential of the VR space and simulated PC in space. \re{These two patterns might help users reduce head rotations, possibly relieving the fatigue for using HMDs~\cite{pavanatto2021we}, though it requires users to put effort into moving the wheeled table. 
}{}

\para{Generalizability}. Considering the fundamental characteristics of our study task, we believe our findings can be generalized to other data-driven sensemaking tasks, especially tasks related to document and network analysis, such as affinity diagramming, social network analysis, and literature review. \re{Moreover, spatial hybrid interfaces can be used for a more general context beyond visual analysis and users, such as note-taking and building knowledge graphs with Large Language Model~\cite{suh2023sensecape} for the general public.}{}

Our findings might work with well-justified 3D visualizations like 3D scatterplots~\cite{yang2020embodied}, 3D heatmaps~\cite{kraus2020assessing}, and space-time cubes~\cite{wagner2019evaluating}. 
Furthermore, similar to separating documents in space, our result suggests using small multiple in VR while complementing an overview in the PC might be effective.

\para{Limitation and Future Work.} Although our results showed positive outcomes and encouraging feedback, there is still room for improvement. While we can estimate usage time for each device using explicit head gaze interaction logs, detailed interactions across devices—such as peeking at another interface—cannot be retrieved.
In future work, eye-tracking data could be collected to understand a more fine-grained usage of different visualizations from different devices.
We did not observe any significant results in performance in these studies, most likely due to the unfamiliarity of our conditions and limited sample size.
In the future, we could recruit more participants with extensive experience in data visualization and conduct a longitudinal study by providing additional usage and training time for each condition, such as text selection in VR and the use of PC and VR devices in PC+VR
We could also recruit more participants to reduce the effect of confounding factors, such as sensemaking skills, reading speed, and limited VR usage experience.
Given its lower physical demand and comparable mental load, we believe the hybrid condition could provide a more sustainable user experience than VR alone. 
Yet, future work still needs to consider how to reduce the physical demand further to the level of PC-only. 
Lastly, future studies could explore the impact of spatial hybrid PC+VR systems with various spatial and temporal settings to enhance visual sensemaking.
Nevertheless, our study provided valuable preliminary results for future research on spatial hybrid interfaces.

\section{Conclusion}
\label{sec:conclusion}

This paper presents a spatial hybrid PC+VR system that seamlessly combines the benefits of both interfaces for visual sensemaking by integrating a simulated PC with a movable mouse and keyboard to support immersive spatial navigation with a VR headset. 
To minimize the effort involved in transitioning between PC and VR, we introduced three techniques: a simulated PC in PC+VR, synchronized states between devices, and hand gestures as the primary modality. 
We conducted a user study with 18 participants to explore user behaviors, usage patterns, and preferences for the PC+VR system compared to PC-only and VR-only conditions during a visual sensemaking task.
\retvcg{We found the following key insights from our exploratory study: 
1) PC+VR was overall most preferred, particularly for interaction; 
2) PC+VR with the movable PC did not negatively impact performance; 
3) PC+VR helped reduce physical demand compared to VR-only; 
4) Participants were more willing to transit between PC and VR when the transition cost was lower, as shown by an increase in temporary transition at the interaction level; and 
5) Participants engaged in more spatial navigation in PC+VR, utilizing features such as moving the simulated PC and rotating around it to enhance their experience.
}{}
We believe our findings could inspire future designs of spatial hybrid systems to enhance visual sensemaking.

\bibliographystyle{abbrv-doi-hyperref}

\bibliography{main}

\begin{IEEEbiography}[{\includegraphics[width=1in,height=1.25in,clip,keepaspectratio]{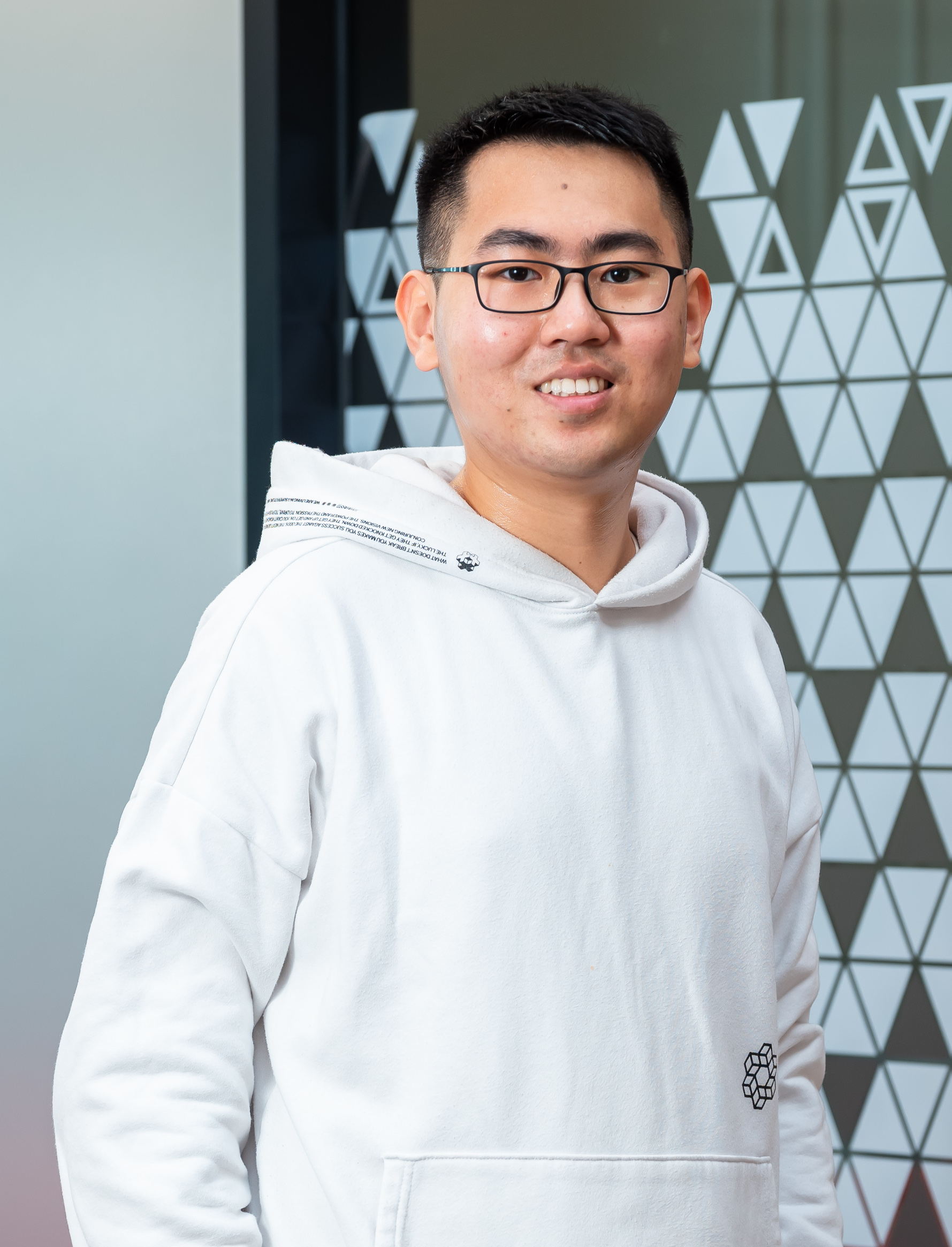}}]{Wai Tong} is an assistant professor in the College of Performance, Visualization and Fine Arts at Texas A\&M University. He obtained his Ph.D. and B.Eng. degrees in the Department of Computer Science and Engineering from the Hong Kong University of Science and Technology. His research interests include data visualization, human-computer interaction, and immersive technology. For more information, please visit \url{https://wtong2017.github.io/}.
\end{IEEEbiography}

\begin{IEEEbiography}[{\includegraphics[width=1in,height=1.25in,clip,keepaspectratio]{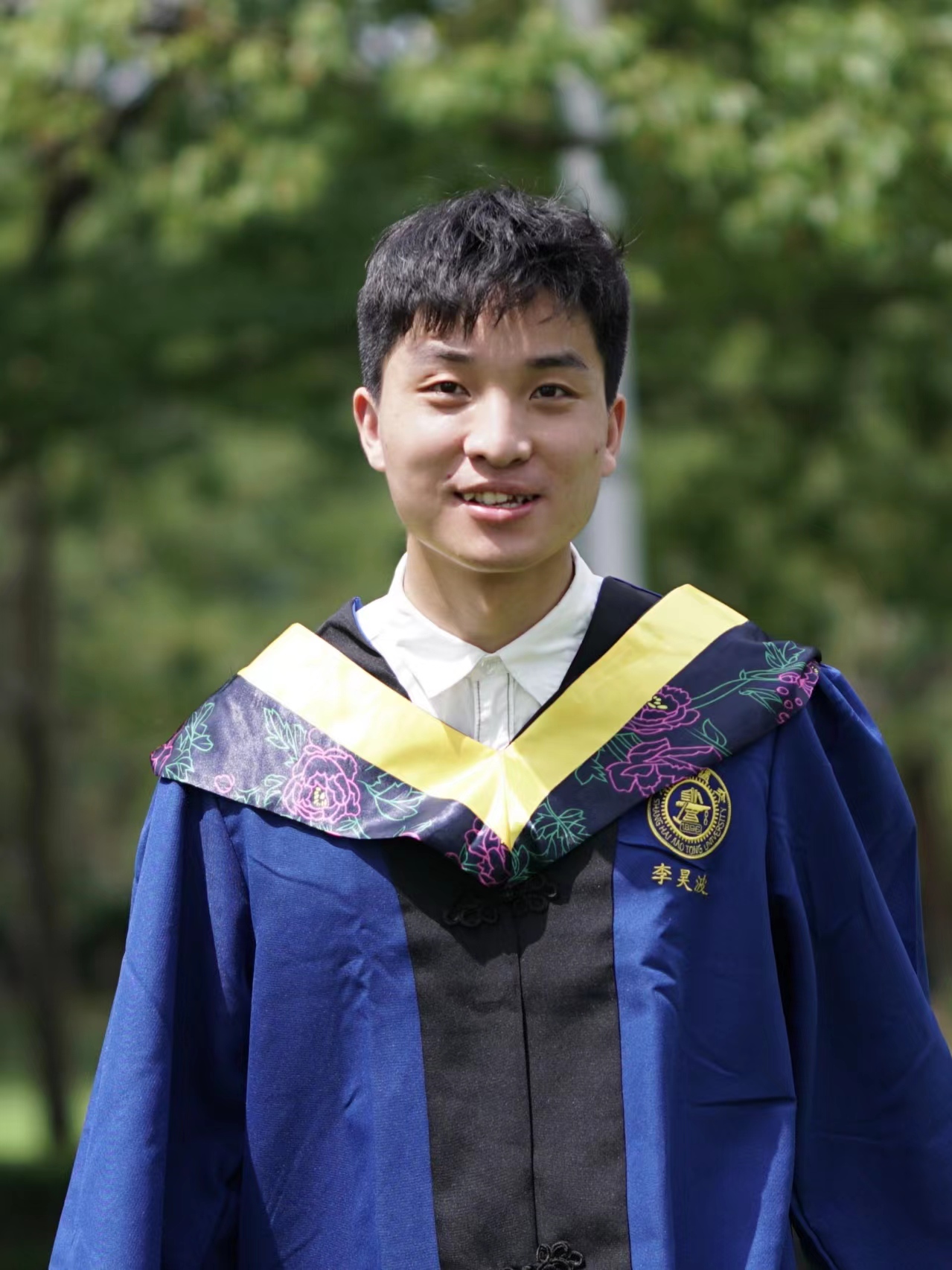}}]{Haobo Li} is a Ph.D. student at HKUST VisLab, Department of Computer Science and Engineering, Hong Kong University of Science and Technology (HKUST). He obtained his master's degree from Shanghai Jiao Tong University. His research interests include data visualization, digital twin, and their applications in the environmental domain. For more information, please visit his website at \url{https://hobolee.github.io/}.
\end{IEEEbiography}

\begin{IEEEbiography}[{\includegraphics[width=1in,height=1.25in,clip,keepaspectratio]{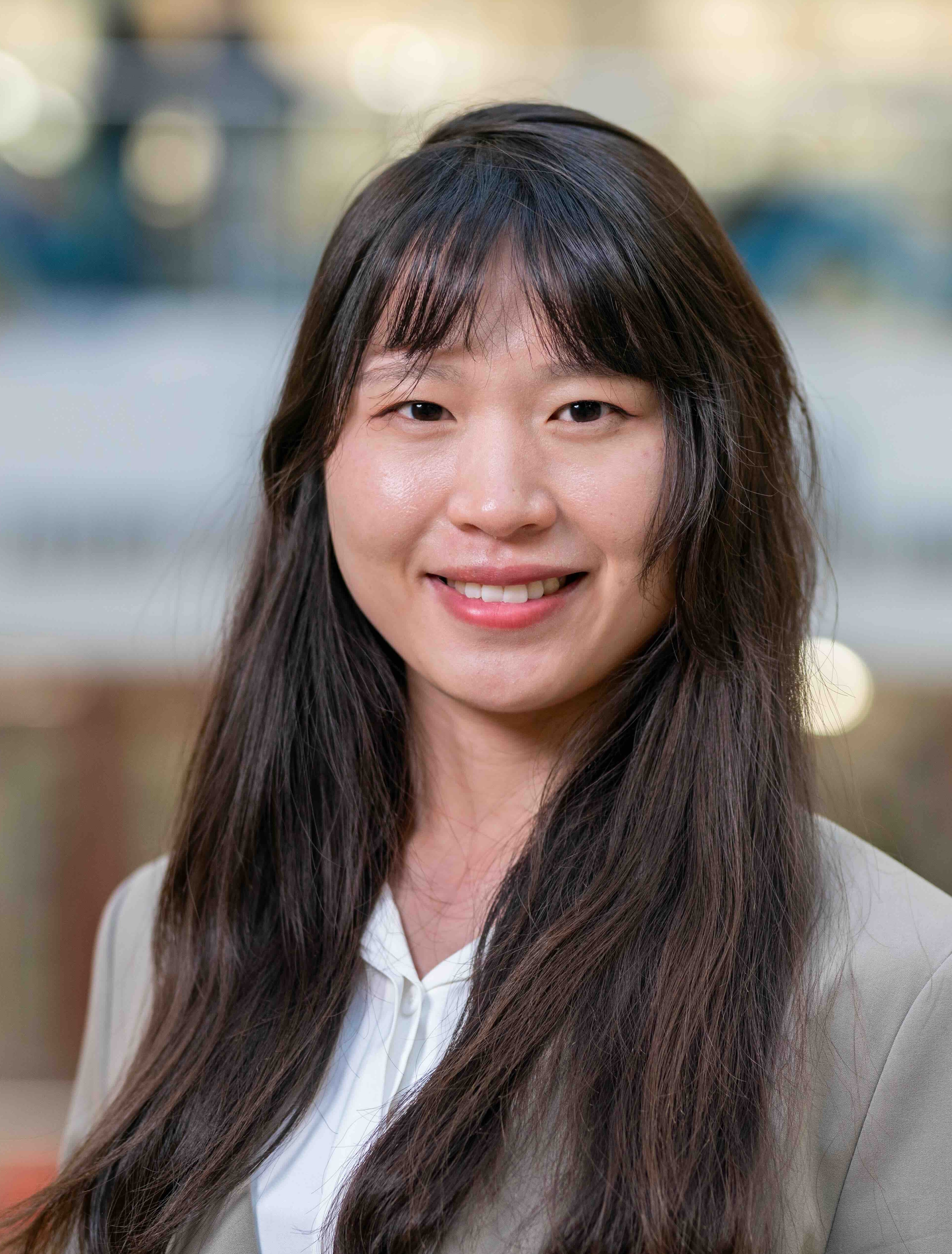}}]{Meng Xia} is an assistant professor in the Department of Computer Science and Engineering at Texas A\&M University. She obtained a Ph.D. degree from the Hong Kong University of Science and Technology and worked as a Postdoc at Carnegie Mellon University and Korea Advanced Institute of Science and Technology, respectively. Her research interests mainly focus on Human-AI Interaction, Data Visualization, and Education Technology. More details can be found at \url{https://www.xiameng.org}.
\end{IEEEbiography}

\begin{IEEEbiography}[{\includegraphics[width=1in,height=1.25in,clip,keepaspectratio]{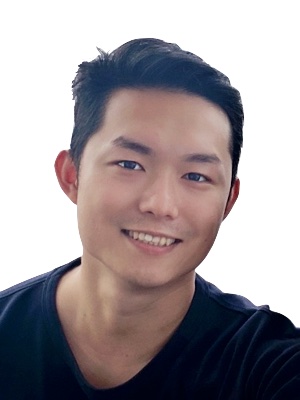}}]{Wong Kam-Kwai} is a Ph.D. candidate in the Department of Computer Science and Engineering at the Hong Kong University of Science and Technology (HKUST). He received his B.E. in HKUST. His main research interests are in data visualization, visual analytics and data mining. For more information, please visit \url{https://kamkwai.com}.
\end{IEEEbiography}

\begin{IEEEbiography}[{\includegraphics[width=1in,height=1.25in,clip,keepaspectratio]{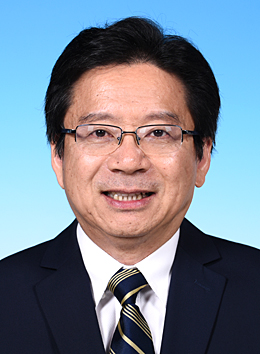}}]{Ting-Chuen Pong}
is a professor at the Department of Computer Science and Engineering at the Hong Kong University of Science
and Technology (HKUST). He received his Ph.D.
in Computer Science from Virginia Polytechnic
Institute and State University in 1984. His research interests include computer vision, multimedia computing, and IT in Education. For more
information, please visit \url{http://www.cse.ust.hk/faculty/tcpong/}.
\end{IEEEbiography}

\begin{IEEEbiography}[{\includegraphics[width=1.0in,height=1.25in,clip,keepaspectratio]{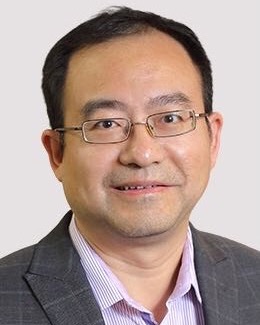}}]{Huamin Qu}
is the dean of the Academy of Interdisciplinary Studies (AIS), the head of the Division of Emerging Interdisciplinary Areas (EMIA), and a chair professor in the Department of Computer Science and Engineering (CSE) at the Hong Kong University of Science and Technology (HKUST), and was the acting head of the Computational Media and Arts Thrust Areas (CMA) at HKUST(GZ). He obtained a BS in Mathematics from Xi'an Jiaotong University, China, an MS and a PhD in Computer Science from the Stony Brook University. His main research interests are in visualization and human-computer interaction, with focuses on urban informatics, social network analysis, E-learning, text visualization, and explainable artificial intelligence (XAI).
For more information, please visit \url{http://huamin.org/}.
\end{IEEEbiography}

\begin{IEEEbiography}[{\includegraphics[width=1in,height=1.25in,clip,keepaspectratio]{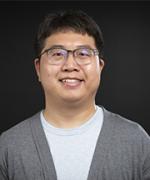}}]{Yalong Yang} is an assistant professor in the School of Interactive Computing at Georgia Institute of Technology.
He was an Assistant Professor at Virginia Tech, a Postdoctoral Fellow at Harvard University, and a Ph.D. student at Monash University, Australia. 
His research designs and evaluates interactive human-data interaction systems on both conventional 2D screens and in 3D immersive environments (VR/AR). For more information, please visit \url{https://ivi.cc.gatech.edu/}.
\end{IEEEbiography}

\vfill

\end{document}